%% file: main.tex
\documentclass[
reprint,
superscriptaddress,
amsmath,amssymb,
aps,
prb,
floatfix,
]{revtex4-2}

\usepackage{graphicx}
\graphicspath{{Figures/}}
\usepackage{dcolumn}
\usepackage{bm}
\usepackage{color}
\usepackage{placeins}
\usepackage[colorlinks=true,
            linkcolor=blue,
            urlcolor=blue,
            citecolor=blue,
            anchorcolor=blue]{hyperref}
\hypersetup{pdftitle={Vibrational Dressing of Collective Molecular Polaritons: A Scalable Green\unichar{8217}s-Function Theory}}
\usepackage{tikz}
\usetikzlibrary{arrows.meta,calc,decorations.markings,positioning,shapes.geometric,decorations.pathmorphing}

\begin{document}
\setcounter{secnumdepth}{0}

\title[]{\large Vibrational Dressing of Collective Molecular Polaritons: A Scalable Green's-Function Theory}

\author{Wenxiang Ying}
\email{wying3@sas.upenn.edu}
\affiliation{Department of Chemistry, University of Pennsylvania, Philadelphia, Pennsylvania 19104, USA}

\author{Michael Galperin}
\email{mgalperin@tauex.tau.ac.il}
\affiliation{School of Chemistry, Tel Aviv University, Tel Aviv 69978, Israel}

\author{Abraham Nitzan}
\email{anitzan@sas.upenn.edu}
\affiliation{Department of Chemistry, University of Pennsylvania, Philadelphia, Pennsylvania 19104, USA}
\affiliation{School of Chemistry, Tel Aviv University, Tel Aviv 69978, Israel}

\begin{abstract}
We develop a scalable Green's-function theory of vibrationally
dressed molecular-polariton absorption in the
Holstein--Tavis--Cummings model.
Within the self-consistent Born approximation (SCBA), the
ground-state linear response of $N$ identical molecules coupled
to one cavity mode follows from a $2\times2$ cavity--bright
matrix, while scattering through the $N-1$ dark states remains
in the molecular self-energy.
The number of self-consistent equations is independent of $N$,
allowing large ensembles to be treated without explicitly
resolving the dark manifold.
Benchmarks against hierarchical equations of motion show that
the theory captures collective linewidth narrowing and an
$N$-dependent minimum in the lower-polariton linewidth as its
exciton fraction varies, beyond a Hopfield-weighted linewidth
estimate.
Comparisons with exact diagonalization for discrete local
vibrations show weaker vibronic progressions and better overall
SCBA agreement as $N$ increases at fixed single-molecule coupling.
The resulting framework connects microscopic vibrational
scattering to collective absorption spectra without introducing
fitted polariton linewidths.
\end{abstract}

\maketitle

Collective coupling of $N$ molecules to a confined optical mode produces upper (UP) and lower (LP) polaritons whose photon and exciton fractions are described by Hopfield coefficients~\cite{tavis1968exact,Tavis_1969}. A common estimate weights the cavity linewidth $\kappa$ and the isolated-molecule linewidth $\Gamma_{\rm ex}$ by these fractions~\cite{HDeng_RMP2010}. This interpolation, however, misses the coherence and scattering generated when site-local molecular environments are expressed in a delocalized collective basis, and cannot account for the linewidth narrowing, the unequal linewidths of UP and LP, and nontrivial detuning dependence observed under strong coupling~\cite{Whittaker_1996, Savona_1997, Ell_1998, Baumberg_1998, Baumberg_1998_2, Borri_2000, Maghrebi_2005, Murphy_2014, Watanabe_2020, Rury_2024, Rury_2024b}. 

The Holstein--Tavis--Cummings (HTC) model provides a microscopic setting for these effects~\cite{Herrera_Spano_PRL,herrera2018theory,Li2021ARPC,Hu_JCP2025,Ying_ARPC2026}. In the pure electronic basis, and in the long-wavelength approximation, a spatially uniform cavity field excites the symmetric combination of molecular transitions (the bright state). 
The other $N-1$ orthogonal molecular combinations have zero matrix element with the cavity field (the dark states). Although these combinations are not eigenstates of the full HTC Hamiltonian once local vibronic coupling is included, they provide a useful basis for resolving the intermediate molecular channels through which local vibrations mediate UP--LP, polariton--dark, and dark--dark scattering~\cite{Chng_2024, Ying_mn24, Lai_JCP2024, Lai_JCP2026, Ying_NatComm2025}, contributing to the bright state self-energy.  

% At the single-emitter level, phonon-dressed cavity QED has been treated microscopically using exact diagonalization and nonequilibrium Green's-function approaches~\cite{Kaer2013,Hornecker2017}.
Collective linewidth narrowing has been described using disorder averaging, multiple scattering, Kubo--Anderson theory, and direct HTC dynamics~\cite{Kinsler_1996,Savona_1997_theory,Climent_2023_kuboanderson,Elious_2023,MondalJCP2025paper1,Ying_mn24}.
A common qualitative picture relates this narrowing to the reduced reorganization energy of the symmetric bright coordinate, $\lambda_{\mathrm{eff}}=\lambda/N$~\cite{Herrera_Spano_PRL,herrera2018theory}. However, this scaling alone does not determine the spectral lineshape: local vibrations also scatter polaritons into the dark manifold, with rates set by the vibrational spectrum, temperature, and polariton--dark energy gaps. A scalable microscopic theory must reduce the optically active problem without discarding the dark manifold or replacing its effects by fitted UP and LP linewidths.

In this Letter, we connect these local scattering processes to collective ground-state absorption with a compact Green's-function theory. The states required for linear absorption are represented exactly, after which the vibrational self-energy is evaluated either once in the Born approximation (BA) or self-consistently in the self-consistent Born approximation (SCBA)~\cite{Fan_1951,Giustino_2017}. We test collective linewidth narrowing, composition-dependent linewidths, and resolved vibronic spectra against hierarchical equations of motion (HEOM)~\cite{Tanimura1990,Xu_2007,Tanimura_2020,Yan_2014}, or mean-field HEOM (MF-HEOM) in the collective limit~\cite{Reichman_Nanoph2024}, and exact diagonalization when appropriate. Formal derivations and numerical checks are given in the Supplemental Material (SM)~\cite{SM}. 

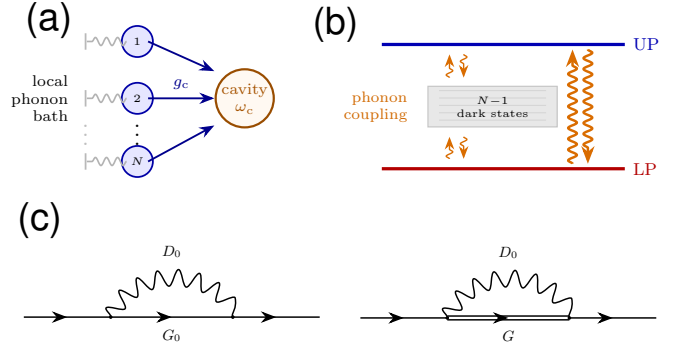
\begin{figure}[tbp]
\centering
\input{Figures/Main/Fig1/Fig1.tex}
\caption{Physical content of the HTC model and its collective scattering channels. (a) Each molecule couples to an independent local phonon bath and uniformly to one cavity mode. (b) Local vibrations connect the bright polaritons to the $N-1$ dark molecular channels and mediate interbranch scattering. (c) The Fan self-energy evaluated with the vibration-free internal propagator $G_0$ gives the single-shot BA (left), whereas replacing it by the dressed propagator $G$ gives the SCBA (right). The phonon line $D_0$ remains bare because $\Pi^R=0$.}
\label{fig:schematic}
\end{figure}

\paragraph{Model---}
We set $\hbar=1$ throughout this paper. The model contains $N$ identical two-level molecules with vertical transition energy $\omega_0$ and raising and lowering operators $\hat\sigma_n^\pm$. The physical cavity operator $\hat a_{\rm ph}$ annihilates a photon of energy $\omega_{\rm c}$, and the uniform single-molecule coupling is $g_{\rm c}$. Each molecule has its own harmonic bath: $\hat b_{\alpha n}$ annihilates a local vibration of frequency $\omega_\alpha$ on molecule $n$, and $c_\alpha$ is its coupling to the molecular excitation. The baths have identical parameters but uncorrelated fluctuations. The Hamiltonian is
\begin{align}
\hat H={}&\omega_0\sum_{n=1}^{N}\hat\sigma_n^+\hat\sigma_n^-
+\omega_{\rm c}\hat a_{\rm ph}^\dagger\hat a_{\rm ph}
+g_{\rm c}\sum_n(\hat a_{\rm ph}^\dagger\hat\sigma_n^-+\hat a_{\rm ph}\hat\sigma_n^+)\notag\\
&+\sum_{n\alpha}\omega_\alpha\hat b_{\alpha n}^\dagger\hat b_{\alpha n}
+\sum_{n\alpha}c_\alpha\hat\sigma_n^+\hat\sigma_n^-(\hat b_{\alpha n}+\hat b_{\alpha n}^\dagger).
\label{eq:HTC}
\end{align}
The spectral density $J(\nu)=\pi\sum_\alpha c_\alpha^2\delta(\nu-\omega_\alpha)$ and reorganization energy $\lambda=\pi^{-1}\int_0^\infty d\nu\,J(\nu)/\nu$ characterize each local bath. The vertical transition energy is $\omega_0=\omega_{00}+\lambda$, where $\omega_{00}$ is the energy difference between the vibrational ground levels of the two electronic states. The rotating-wave approximation makes the total molecular-plus-cavity excitation number of the closed Hamiltonian conserved, while the long-wavelength approximation gives the same $g_{\rm c}$ on every molecule. There is no direct electronic coupling between molecules. The cavity mediates excitation transfer between sites, while local vibrations scatter excitations between the collective bright and dark molecular combinations (Fig.~\ref{fig:schematic}a--b).

Photons also escape into external far-field radiation modes. We take the leakage rate to be constant over the frequencies of interest and include it through the photon self-energy $\Sigma_{\rm ph}^R=-i\kappa/2$, with a constant real frequency shift absorbed into $\omega_{\rm c}$. Thus $\kappa$ is the cavity spectral full width at half maximum (FWHM). % and $\kappa/2$ is the field-amplitude decay rate. 
This loss is included separately from the excitation-number conservation of the closed Hamiltonian.

For linear absorption from the electronic--photonic ground state $|G\rangle$, the dipole operator creates one excitation. Intermediate states contain either one cavity photon, $|C\rangle=\hat a_{\rm ph}^\dagger|G\rangle$, or one excited molecule, $|n\rangle=\hat\sigma_n^+|G\rangle$, while the vibrations may contain any number of quanta. We introduce canonical bosonic operators $\hat a$ and $\hat d_n$, with their vacuum $|\mathrm{vac}\rangle$ representing the physical electronic--photonic ground state $|G\rangle$. The one-excitation states are represented as $|C\rangle\leftrightarrow\hat a^\dagger|\mathrm{vac}\rangle$ and $|n\rangle\leftrightarrow\hat d_n^\dagger|\mathrm{vac}\rangle$. The Hamiltonian in this representation is
\begin{subequations}
\begin{align}
\hat{\widetilde H}={}&\omega_{\rm c}\hat a^\dagger\hat a
+\omega_0\sum_n\hat d_n^\dagger\hat d_n
+g_{\rm c}\sum_n(\hat a^\dagger\hat d_n+\hat d_n^\dagger\hat a)\notag\\
&+\sum_{n\alpha}\omega_\alpha\hat b_{\alpha n}^\dagger\hat b_{\alpha n}
+\sum_{n\alpha}c_\alpha\hat d_n^\dagger\hat d_n(\hat b_{\alpha n}+\hat b_{\alpha n}^\dagger), \label{eq:reduced_H}\\[-2pt]
&\hat Q_{\rm exc}\equiv\hat a^\dagger\hat a+\sum_n\hat d_n^\dagger\hat d_n\in\{0,1\}.
\end{align}
\end{subequations}
See details in SM~\cite{SM} Sec.~II\,A. Within $\hat Q_{\rm exc}\in\{0,1\}$, Eq.~\ref{eq:reduced_H} has the same matrix elements as the original Hamiltonian: one optical creation therefore takes $Q_{\rm exc}$ from 0 ($|\mathrm{vac}\rangle$) to 1, subsequent evolution under the closed Hamiltonian between the optical vertices remains in that subspace, and the final annihilation returns the state to the vacuum. This representation is exact for ground-state linear absorption. The approximation is introduced only later, when the effect of the vibrations is evaluated with BA or SCBA. Extensions to nonlinear response or initially excited populations require the relevant occupations and excitation manifolds to be treated explicitly.

\paragraph{Green's function formalism---}
Before the probe is applied, the molecules are electronically unexcited, the cavity is empty, and the vibrations are in thermal equilibrium at inverse temperature $\beta=(k_{\rm B}T)^{-1}$. The corresponding initial density operator is $\hat\rho_0=|{\rm vac}\rangle\langle{\rm vac}|\otimes\hat\rho_{\rm B}^{\rm th}$, with $\hat\rho_{\rm B}^{\rm th}$ the thermal state of the ground-state vibrational baths.
The Green's function is
\begin{align}
G_{ij}^R(t)&=-i\theta(t)\operatorname{Tr}\!\left\{\hat\rho_0
[\hat q_i(t),\hat q_j^\dagger(0)]\right\},
\label{eq:retarded_GF}
\end{align}
where $\hat{\boldsymbol q}=(\hat a,\hat d_1,\ldots,\hat d_N)^T$, and $i,j\in\{0,1,\ldots,N\}$, with $\hat q_0 \equiv \hat a$. The step function $\theta(t)$ is zero for $t<0$ and one for $t>0$, so the superscript $R$ denotes a response that vanishes before the excitation is applied. Specifically, $\hat q_j^\dagger$ introduces an excitation in component $j$ at time zero, and $G_{ij}^R$ is the amplitude detected in component $i$ later, averaged over the initial thermal vibrations. We use the Fourier convention $G^R(\omega)=\int_{-\infty}^{\infty}dt\,e^{i\omega t}G^R(t)$.

Before using molecular symmetry, $\mathbf G^R$ is an $(N+1)\times(N+1)$ site-basis matrix. We define $\mathbf G_0^R$ by switching off the vibronic coupling while retaining both the light--matter coupling and cavity leakage. 
Because the Holstein vibronic coupling is a sum over molecules, each coupled to an independent bath with the same spectral density, the site-local Fan closure used in BA and SCBA gives a uniform site-diagonal self-energy matrix, $\Sigma_{nm}^R(\omega)=\delta_{nm}\Sigma^R(\omega)$, and thereby enables the bright--dark reduction below. Here, $\Sigma^R(\omega)$ is the Fan self-energy~\cite{Fan_1951, Giustino_2017} of a molecule interacting with its own phonon bath, whose real and imaginary parts shift and damp the molecular response. The resulting Dyson equation is
\begin{equation}
[\mathbf G^R(\omega)]^{-1}=[\mathbf G_0^R(\omega)]^{-1}
-\operatorname{diag}\!\left(0,\Sigma^R(\omega)\mathbf 1_N\right),
\label{eq:full_Dyson}
\end{equation}
where $\mathbf 1_N$ is the identity in the molecular site subspace. 

We further define the collective molecular modes
$\hat d_k=N^{-1/2}\sum_{n=1}^{N}
e^{-2\pi i kn/N}\hat d_n$, with $k=0,1,\ldots,N-1$.
The $k=0$ mode is bright, $\hat d_0\equiv\hat B$,
whereas the remaining modes are dark.
% Similarly, define the bath modes $\hat b_{\alpha q}=N^{-1/2}\sum_{n=1}^{N} e^{-iqn}\hat b_{\alpha n}$, where $q$ takes the same discrete momenta.
Only $\hat B$ couples to the cavity, with matrix element $\sqrt N g_{\rm c}$. In this basis, $\Sigma_{kk'}^R(\omega)=\delta_{kk'}\Sigma^R(\omega)$. The full inverse Green's function is the direct sum
\begin{equation}
[\mathbf G^R(\omega)]^{-1}
=
[\mathbf G_{aB}^R(\omega)]^{-1}
\bigoplus_{k\neq0}
\left[G_{D_k}^R(\omega)\right]^{-1}.
\label{eq:directsum}
\end{equation}
where the optically active block is
{\small
\begin{equation}
\left[\mathbf G_{aB}^R(\omega)\right]^{-1}=
\begin{pmatrix}
\omega-\omega_{\rm c}+i\kappa/2 & -\sqrt N g_{\rm c}\\
-\sqrt N g_{\rm c} & \omega-\omega_0+i0^+-\Sigma^R(\omega)
\end{pmatrix}.
\label{eq:brightblock}
\end{equation}
}
Here, we introduce $G^R_{BB}$ as the bright--bright matrix element. Inverting this matrix gives the Green's function of the bright molecular excitation,
\begin{equation}
G_{BB}^R(\omega)=\frac{1}{\omega-\omega_0+i0^+
-\frac{Ng_{\rm c}^2}{\omega-\omega_{\rm c}+i\kappa/2}-\Sigma^R(\omega)}.
\label{eq:bright_schur}
\end{equation}
When $\kappa=0$, the photon denominator has the same infinitesimal $+i0^+$ prescription. For later reference, the vibration-free polariton splitting is $\Omega_{\rm R}=(\Delta^2+4Ng_{\rm c}^2)^{1/2}$, with detuning $\Delta=\omega_{\rm c}-\omega_0$. 
Meanwhile, the $N-1$ dark modes form degenerate scalar blocks in Eq.~\ref{eq:directsum}, with
\begin{equation}
G_D^R(\omega) \equiv G_{D_k}^R(\omega) =
[\omega-\omega_0+i0^+-\Sigma^R(\omega)]^{-1}.
\label{eq:GD}
\end{equation}
The resulting diagonal form in the collective basis does not imply the absence of phonon-mediated bright--dark scattering, because the internal line is summed over all collective molecular modes. 
The Green's function for an excitation created and removed on the same molecule (that enters the Fan self-energy, see Eq.~\ref{eq:SCBA} below) is
\begin{equation}
\bar G^R(\omega)\equiv G_{nn}^R(\omega)
=\frac{G_{BB}^R(\omega)+(N-1)G_D^R(\omega)}{N}.
\label{eq:Gbar}
\end{equation}
One sees the dark modes with multiplicity of $N-1$ remain incorporated in $\Sigma^R$ through the contribution $(N-1)G_D^R/N$. 
This dimensional reduction therefore preserves the dark-state contribution to phonon scattering. 
The full Hopfield transformation and polariton-basis Green's functions are given in Sec.~II-D of the SM~\cite{SM}.

\paragraph{Self-consistent Born approximation---} 
The Fan self-energy~\cite{Fan_1951, Giustino_2017} contains one molecular propagator and one phonon propagator, as shown schematically in Fig.~\ref{fig:schematic}c. 
In the zero-excitation reference state,
$\hat q_i|\mathrm{vac}\rangle=0$, the phonon self-energy $\Pi^R$ 
%and the Hartree term $\langle \hat{X}_n \rangle \equiv \langle \sum_{\alpha}c_\alpha(\hat{b}_{\alpha,n}+\hat{b}^\dagger_{\alpha,n}) \rangle$
vanishes, so the phonon propagator remains bare, $D^R=D_0^R$. 
The full phonon Green's-function derivation is given in Sec.~II-C of the SM~\cite{SM}.
Using the spectral-density convention above, the self-consistent Born equation is
\begin{align}
\Sigma^R(\omega)=\frac{1}{\pi}\int_0^\infty d\nu\,J(\nu)
\bigl\{&[n_{\rm B}(\nu)+1]\bar G^R(\omega-\nu)\notag\\[-2pt]
&+n_{\rm B}(\nu)\bar G^R(\omega+\nu)\bigr\},
\label{eq:SCBA}
\end{align}
where $n_{\rm B}(\nu)=(e^{\beta\nu}-1)^{-1}$. The first term describes emission of a vibration and samples the molecular response at $\omega-\nu$; the second describes absorption of a thermally occupied vibration and samples it at $\omega+\nu$. 
Eqs.~\ref{eq:brightblock}, \ref{eq:Gbar}, and \ref{eq:SCBA} are solved together. Starting from a trial $\Sigma^R$, Eqs.~\ref{eq:brightblock} and \ref{eq:Gbar} determine the optical bright and site-local Green's functions; Eq.~\ref{eq:SCBA} then updates $\Sigma^R$. The cycle is repeated until numerical convergence, and the final $G_{BB}^R$ gives the absorption through Eq.~\ref{eq:absorption} below. The closure involves a fixed number of frequency-dependent functions, independent of $N$, while retaining the dark-state contribution through Eq.~\ref{eq:Gbar}. 
The numerical convolution, mixing, and convergence procedures are described in Sec.~IV-B of the SM~\cite{SM}.
In the single-shot Born approximation (BA), Eq.~\ref{eq:SCBA} is instead evaluated once with the vibration-free $\bar G_0^R$. 
Self-consistency sums the noncrossing rainbow diagrams built from this Fan term. It does not include crossed diagrams or vertex corrections, and therefore cannot reproduce the full multiphonon Franck--Condon structure in the strong exciton--phonon coupling regime. See Sec.~II-B of the SM~\cite{SM} for the diagrammatic expansion.

\begin{figure*}[t!]
\centering
\includegraphics[width=0.98\linewidth]{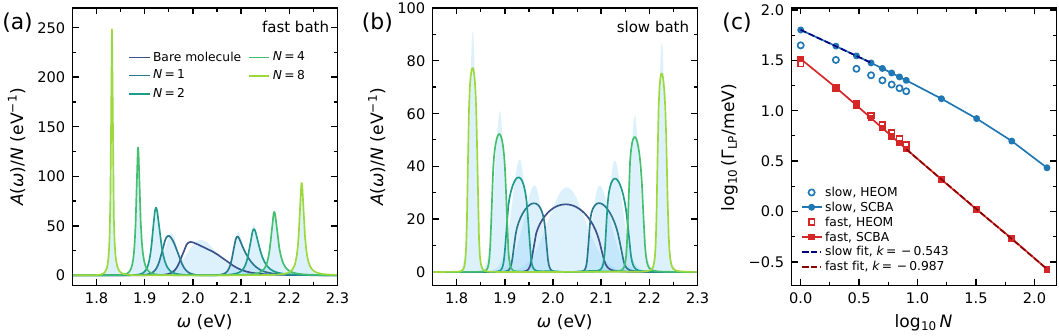}
\caption{Collective linewidth narrowing at vertical resonance ($\omega_{\rm c}=\omega_0=2.03\,\mathrm{eV}$, $\kappa=0$). (a,b) Molecular absorption per molecule for the bare molecule and $N=1,2,4,8$ at fixed $g_{\rm c}=68.1\,\mathrm{meV}$ for the fast and slow baths, respectively: SCBA (solid curves) and HEOM (blue shading). (c) LP FWHM at fixed $\Omega_{\rm R}=385.2\,\mathrm{meV}$: SCBA (solid lines) and HEOM (open points). For both SCBA curves in panel (c), the numerical retarded half-width is $\eta=0.01\,\mathrm{meV}$ for $N\leq64$ and $0.001\,\mathrm{meV}$ for $N=128$. HEOM data are reproduced from Ref.~\citenum{Ying_mn24}.}
\label{fig:linewidth_narrowing}
\end{figure*}

\paragraph{Polariton linear absorption spectra---}
% We are interested in calculating the polariton linear absorption spectra. 
In the long--wavelength limit, that is,
for a uniform molecular optical probe, the transition operator couples to the collective bright mode $\hat B$ since $\hat\mu^+=\mu_0\sum_n\hat d_n^\dagger=\mu_0\sqrt N\,\hat B^\dagger$, where $\mu_0$ is the transition dipole amplitude of one molecule. % As a result, the dipole-dipole response function $\chi^R_{\mu  \mu}(t) = N |\mu_0|^2 G_{BB}^R(t)$, and the absorption lineshape is
Retaining the positive-frequency resonant part of the dipole response gives the absorption lineshape
\begin{align}
A(\omega)&\propto-N|\mu_0|^2\operatorname{Im}G_{BB}^R(\omega).
\label{eq:absorption}
\end{align}
Note that this is a dipole-absorption spectrum, not a port-resolved transmission or reflection signal. 
% The operator $\hat B^\dagger$ need not create an eigenstate: the Lehmann representation of $G_{BB}^R$ sums transitions from the zero-excitation thermal ensemble to all exact one-excitation vibronic eigenstates, weighted by their bright-state matrix elements.
We further note that one may couple the incident far field to the cavity mode, which is often the more natural description of a cavity experiment~\cite{Aizpurua_acsnano2016}. For the absorption spectrum considered here, the two probe models yield similar lineshape, differing only by spectral weights, although absolute intensities and other observables may depend on the specific drive and detection model.
% Any display normalization is stated in the figure captions.

\paragraph{Collective linewidth narrowing---}
For the continuous environment we use the Drude--Lorentz form $J(\nu)=2\lambda\omega_f\nu/(\omega_f^2+\nu^2)$ with the reorganization energy $\lambda=30\,\mathrm{meV}$ and $T=300\,\mathrm{K}$. Unless stated otherwise, we take $\omega_{00}=2.00\,\mathrm{eV}$, $\omega_0=2.03\,\mathrm{eV}$, and $\omega_{\rm c}=\omega_0$. The bath characteristic frequency is $\omega_f$. Following Ref.~\citenum{Ying_mn24}, we compare a slow bath with $\omega_f=2.48\,\mathrm{meV}$ and a fast bath with $\omega_f=24.8\,\mathrm{meV}$, meaning the ratio between the bath correlation rate and the amplitude of the energy-gap fluctuations. 
In the classical high-temperature limit ($k_\mathrm{B}T \gg \omega_f$), 
the energy-gap fluctuation experienced by the bright state,
$\hat X_B=N^{-1}\sum_n\hat X_n$ (where $\hat{X}_n = \sum_\alpha c_\alpha (\hat{b}_{\alpha,n} + \hat{b}^\dagger_{\alpha,n})$), has the correlation function $\operatorname{Re} C_B(t) = 2 \lambda_{\rm eff}\, k_\mathrm{B} T e^{- \omega_f |t|}$, where $C_B(t) \equiv \langle \hat{X}_B(t) \hat{X}_B(0) \rangle$. Its root-mean-square fluctuation amplitude is therefore $\Delta_N = \sqrt{C_B(0)} = \sqrt{2 \lambda_{\rm eff}\, k_\mathrm{B} T}$.
When $k_\mathrm{B}T \gg \omega_f$ and $\Omega_\mathrm{R} \gg \omega_f$, we introduce~\cite{Ying_mn24} $\mathcal K_N \equiv \omega_f / \Delta_N$ as a regime indicator: $\mathcal K_N \ll 1$ indicates the slow modulation (static, inhomogeneous) limit, while $\mathcal K_N \gg 1$ defines the fast modulation (homogeneous) limit~\cite{Mukamel}. Note that this classical estimate serves only as a regime indicator; the full polariton self-energy also contains dark-state and interbranch contributions. $\mathcal K_N$ alone therefore does not determine the linewidth, especially outside the regime $\Omega_\mathrm{R} \gg \omega_f$.

Figure~\ref{fig:linewidth_narrowing} uses two complementary protocols without cavity loss. Panels (a,b) hold $g_{\rm c}=68.1\,\mathrm{meV}$ fixed, so the collective splitting grows as $\sqrt N$. Panel (c) instead holds the resonant Rabi splitting at $\Omega_{\rm R}=385.2\,\mathrm{meV}$, so $g_{\rm c}\propto N^{-1/2}$ and the principal polariton energy scale remains fixed. 
SCBA and HEOM both show progressive narrowing with $N$, and the fast bath narrows more rapidly. Over the fitting intervals specified in SM, Sec.~IV\,F~\cite{SM}, the SCBA results give $\Gamma_{\rm LP}\propto N^{-0.543}$ for the slow bath at $N=1$--4 and $N^{-0.987}$ for the fast bath at $N=8$--128, close to the familiar $N^{-1/2}$ and $N^{-1}$ trends for slow and fast modulation~\cite{Ying_mn24}. 
Nevertheless, to the best of our knowledge, these distinct $N$-scaling laws have not yet been experimentally verified in molecular polariton systems. Experimental tests of these distinct scaling regimes would therefore be valuable.

\begin{figure}[t!]
\centering
\includegraphics[width=0.98\columnwidth]{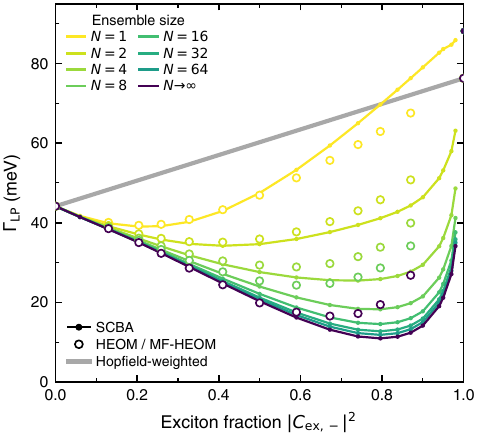}
\caption{LP FWHM versus the (vibration-free) exciton fraction for independent Drude--Lorentz baths ($\lambda=30\,\mathrm{meV}$, $\omega_f=24.8\,\mathrm{meV}$, $T=300\,\mathrm{K}$) and cavity FWHM $\kappa=44.15\,\mathrm{meV}$. Detuning is varied at fixed $\sqrt N g_{\rm c}=136.2\,\mathrm{meV}$. Solid curves are SCBA; open points are HEOM for $N\leq8$ and MF-HEOM for $N\to\infty$. The gray line is Eq.~\ref{eq:hopfield_width} with the HEOM molecular linewidth $\Gamma_{\rm ex}=76.28\,\mathrm{meV}$; the corresponding SCBA linewidth for bare molecule is $88.15\,\mathrm{meV}$. Small-$N$ HEOM data are reproduced from Ref.~\citenum{Ying_mn24}.}
\label{fig:detuning}
\end{figure}

\paragraph{Detuning dependence---}
Figure~\ref{fig:detuning} shows the LP linewidth as detuning is varied at fixed collective coupling $\sqrt N g_{\rm c}=136.2\,\mathrm{meV}$. The horizontal axis is the exciton fraction of LP without vibronic coupling and cavity loss, $|C_{{\rm ex},-}|^2=[1+\Delta/\Omega_{\rm R}]/2$; the corresponding UP exciton fraction is $|C_{{\rm ex},+}|^2=1-|C_{{\rm ex},-}|^2$. These quantities label the bare cavity--bright eigenvectors rather than fitted compositions of the vibronic peaks. The SCBA curves fall below the pure-cavity width, pass through an $N$-dependent minimum, and then rise toward the endpoint of bare molecular linewidth. For the finite ensemble sizes shown, increasing $N$ deepens the minimum and moves it toward a more excitonic LP.

The curves and their $N$-dependence contrast directly with the composition-weighted estimate
\begin{equation}
\Gamma_\pm^{\rm H}=|C_{{\rm ph},\pm}|^2\kappa+|C_{{\rm ex},\pm}|^2\Gamma_{\rm ex},
\label{eq:hopfield_width}
\end{equation}
which is a straight line between bare cavity and molecular linewidths and has no explicit $N$ dependence at fixed composition. Here the gray curve in Fig.~\ref{fig:detuning} uses the bare molecular FWHM $\Gamma_{\rm ex}=76.28\,\mathrm{meV}$ from HEOM, while the SCBA curves approach their separately calculated bare molecular linewidth $88.15\,\mathrm{meV}$ as $|C_{{\rm ex},-}|^2$ approaches unity. The HEOM results for $N=1,2,4,8$ and the MF-HEOM results in the collective limit provide independent benchmarks across the scan of $|C_{{\rm ex},-}|^2$. Additional spectra and the UP FWHM are shown in SM~\cite{SM} Fig.~S10. In that comparison, SCBA overestimates the linewidth of a strongly excitonic UP. 
%; SM Sec.~V\,B analyzes the contribution from the nearby dark manifold.

\begin{figure*}[t!]
\centering
\includegraphics[width=0.98\linewidth]{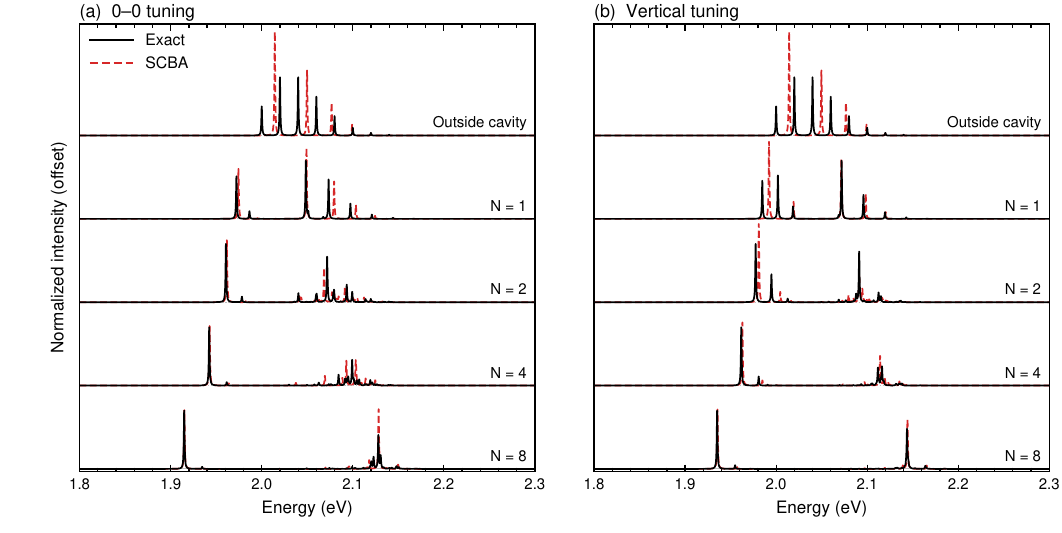}
\caption{Vibronic absorption spectra for one vibrational mode per molecule
at $S=2$: exact (black solid) and SCBA (red dashed).
The cavity is tuned to (a) the 0--0 transition,
$\omega_{\rm c}=\omega_{00}=2.00\,\mathrm{eV}$, or
(b) the vertical transition,
$\omega_{\rm c}=\omega_0=2.04\,\mathrm{eV}$.
From top to bottom, each panel shows the outside-cavity spectrum
and cavity spectra for $N=1,2,4,8$ at fixed
$g_{\rm c}=35.36\,\mathrm{meV}$.
Each SCBA spectrum is area-matched to its exact counterpart over
$1.8$--$2.3\,\mathrm{eV}$. Each pair is then divided by the exact peak
height and vertically offset.
Other parameters are $\nu_0=20\,\mathrm{meV}$,
$\lambda=40\,\mathrm{meV}$, $T=0$, and
Lorentzian half-width $\eta=0.5\,\mathrm{meV}$.}
\label{fig:vibronic}
\end{figure*}

\paragraph{Vibronic spectra---}
The continuous spectral densities used in
Figs.~\ref{fig:linewidth_narrowing}--\ref{fig:detuning}
emphasize spectral envelopes and linewidths relevant to condensed-phase
systems. We now consider one independent vibrational mode per molecule
to resolve discrete vibronic structure, motivated by the recent gas-phase
polariton experiments~\cite{Wright2023Methane,Nelson2026Iodine} and simulations~\cite{Hoffmann_PRL2025, bustamante_2026, Ying_JCTC2026, Ying_JPCL2026}.

Figure~\ref{fig:vibronic} compares 0--0 tuning, $\omega_{\rm c}=\omega_{00}$, with vertical-transition tuning, $\omega_{\rm c}=\omega_0$. The outside-cavity Franck--Condon profile provides the same reference in both panels. Here we fix the vibronic coupling strength corresponding to a Huang--Rhys factor $S \equiv \lambda / \nu_0 = 2$, as well as $g_{\rm c}=35.36\,\mathrm{meV}$. We also adopt a Lorentzian half-width $\eta=0.5\,\mathrm{meV}$, corresponding to the numerical value of positive infinitesimal $0^+$.
Because $g_{\rm c}$ is fixed, increasing $N$ from 1 to 8 raises $\sqrt N g_{\rm c}$ from $35.36$ to $100.01\,\mathrm{meV}$ and changes the energy separation of the collective states as well as their vibrational dressing.
For both tuning choices, the vibronic progression becomes less prominent
as $N$ increases at fixed $g_c$, and spectral weight becomes more
concentrated around the dominant polariton peaks. This sharpening of
the vibronic envelopes is qualitatively consistent with the collective
narrowing in Fig.~\ref{fig:linewidth_narrowing}. SCBA also shows better
overall agreement with the exact spectra at larger $N$, particularly
for the dominant LP peak, although discrepancies remain in the
UP-region multiphonon structure and relative peak intensities at $S=2$.
SM~\cite{SM} Figs.~S4 and S5 provide the full $N$ and $S$ dependence
and numerical checks.

\FloatBarrier

\paragraph{Conclusions---}
We have developed a scalable Green's-function theory of vibrational
dressing in collective molecular-polariton absorption.
For identical molecules coupled uniformly to one cavity mode,
SCBA reduces the optical calculation to a $2\times2$ cavity--bright
matrix while retaining dark-state scattering in the molecular
self-energy. The resulting self-consistent equations do not grow
in number with $N$.
Comparisons with HEOM show that the theory captures distinct
collective linewidth-narrowing trends for slow and fast baths,
as well as an $N$-dependent minimum in the LP linewidth versus
exciton fraction that is absent from a Hopfield-weighted estimate.
For discrete local vibrations, increasing $N$ at fixed $g_{\rm c}$
weakens the vibronic progression and concentrates spectral weight
around the dominant polariton peaks, with better overall agreement
between SCBA and exact diagonalization.
These results connect local vibrational scattering to collective
spectral narrowing without introducing fitted polariton linewidths.
Quantitative discrepancies remain for strongly excitonic UP
linewidths and resolved multiphonon structure.
Several directions remain open for future work, including
treatments beyond SCBA~\cite{Silbey_JCP1977, Silbey_JCP1980, Silbey_JCP2008, Segal_2023, Segal_2024, Segal_2025}, multimode cavities~\cite{Ying2023NanoP, Krauss_ACSNano2024, Krauss_JCP2024, Mandal_NL2023, Chng_NL2025, Mandal_PRB2025, Mandal_ACSPh2025, Mandal_JPCL2026}, and nonlinear spectroscopy~\cite{Galperin_JCP2025, Galperin_JPCL2025, Galperin_PRR2025, Yuen-Zhou_PRL2025}.\\

\paragraph{Acknowledgments---}
This work was supported by the European Research Council under ERC-2024-SyG-101167294; UnMySt.\\

\paragraph{Data availability---}
The data and figure-reproduction code that support the findings of this work
are available at
\url{https://github.com/Okita0512/Polariton_Spectra_GF}.

\nocite{Galperin_2015,Hornecker2017,Kaer2013,LDEOM_2023,Nitzan,Pade_2010,Pade_2011,Tanimura_2006,Wang_Thoss_2006,Wang_Thoss_2007,White_2012,White_2014,Xu_2005,Yan_2016}
\bibliography{Ref}

\end{document}

\raggedbottom

\title[]{\texorpdfstring{\large Supplemental Material for \\ Vibrational Dressing of Collective Molecular Polaritons: A Scalable Green's-Function Theory}{Supplemental Material for Vibrational Dressing of Collective Molecular Polaritons: A Scalable Green's-Function Theory}}

\author{Wenxiang Ying}
\email{wying3@sas.upenn.edu}
\affiliation{Department of Chemistry, University of Pennsylvania, Philadelphia, Pennsylvania 19104, USA}

\author{Michael Galperin}
\email{mgalperin@tauex.tau.ac.il}
\affiliation{School of Chemistry, Tel Aviv University, Tel Aviv 69978, Israel}

\author{Abraham Nitzan}
\email{anitzan@sas.upenn.edu}
\affiliation{Department of Chemistry, University of Pennsylvania, Philadelphia, Pennsylvania 19104, USA}
\affiliation{School of Chemistry, Tel Aviv University, Tel Aviv 69978, Israel}

\maketitle
\hypersetup{pdftitle={Supplemental Material for Vibrational Dressing of Collective Molecular Polaritons: A Scalable Green's-Function Theory}}
\setcounter{tocdepth}{2}
\begingroup
%\footnotesize
\tableofcontents
\endgroup

\clearpage
%%%%%%%%%%%%%%%%%%%%%%%%%%%%%%%%%%%%%%%%%%
\section{HEOM and MF-HEOM Methods}\label{sec:SI_HEOM_details}

\subsection{HEOM with Cavity Loss}\label{sec:I-A}

We treat the molecular phonon baths with the hierarchical equations of motion (HEOM)~\cite{Tanimura1990, Tanimura_2006, Xu_2005, Xu_2007} and cavity loss with a Lindblad dissipator~\cite{LDEOM_2023}.
The equations used here are summarized below; derivations are given in Refs.~\citenum{Yan_2014, Yan_2016, LDEOM_2023}.

The total system--bath Hamiltonian is
\begin{align}
    \hat{H} = \hat{H}_{\mathrm{S}} + \hat{h}_{\textsc{B}} + \hat{H}_{\mathrm{SB}},
    \qquad
    \hat{H}_{\mathrm{SB}} = \sum_{a = 1}^\mathcal{M} \hat{Q}_{a} \otimes \hat{F}_a,
\end{align}
where $\hat{H}_{\mathrm{SB}}$ describes linear system-bath coupling with $\mathcal{M}$ modes and the reorganization shift is included directly in $\hat H_\mathrm{S}$, as in the main text.
The bath force operators $\hat{F}_a$ are decomposed into dissipaton components~\cite{Yan_2014, Yan_2016},
\begin{align}
    \hat{F}_a = \sum_{k = 1}^{K} \hat{f}_{ak},
\end{align}
with bath time-correlation function
\begin{align} \label{bare_bath_TCF}
    C_{ab}(t) \equiv \langle \hat{F}_a(t) \hat{F}_b(0) \rangle_{\textsc{B}} = \sum_{k = 1}^{K} \eta_{abk} e^{- \gamma_{ak} t}.
\end{align}
The dynamical variables in HEOM are the dissipaton density operators (DDOs),
\begin{align}\label{DDO}
    \rho^{(n)}_{\mathbf{n}} (t) \equiv \mathrm{Tr}_{\textsc{B}} \left[ \Big( \prod_{ak} \hat{f}^{n_{ak}}_{ak}  \Big)^\circ \hat{\rho}_{\mathrm{T}} (t) \right],
\end{align}
where $(\cdots)^\circ$ denotes irreducible ordering, $\mathbf{n} \equiv \{\cdots, n_{ak}, \cdots\}$ is the tier index with $n = \sum_{ak} n_{ak}$, and $\hat{\rho}_{\mathrm{S}}(t) \equiv \rho^{(0)}_{\mathbf{0}}(t)$ is the reduced system density operator.

The mixed HEOM-Lindblad equations of motion are~\cite{LDEOM_2023}
\begin{align} \label{eq-LDEOM}
    \dot{\rho}^{(n)}_{\bm{\mathrm{n}}}(t) =& - \left( i \mathcal{L}_{\mathrm{S}} - \mathcal{L}_\kappa + \sum_{ak} n_{ak} \gamma_{ak} \right) \rho^{(n)}_{\bm{\mathrm{n}}}(t)
    - i \sum_{ak} \mathcal{Q}^{\times}_a \rho^{(n + 1)}_{\bm{\mathrm{n}}^+_{ak}}(t) \notag \\
    &- i \sum_{abk} n_{ak} \left( \eta'_{abk} \mathcal{Q}^{\times}_b + i \eta^{''}_{abk} \mathcal{Q}^{\circ}_b \right)\rho^{(n-1)}_{\bm{\mathrm{n}}^{-}_{ak}} (t),
\end{align}
with superoperators and coefficients
\begin{align}
    \mathcal{L}_{\mathrm{S}} \hat{O} &\equiv [\hat{H}_{\mathrm{S}}, \hat{O}], \quad
    \mathcal{Q}^{\times}_a \hat{O}  \equiv [\hat{Q}_{a}, \hat{O}], \quad
    \mathcal{Q}^{\circ}_a \hat{O} \equiv \{ \hat{Q}_{a}, \hat{O} \}, \notag\\
    \eta'_{abk} &\equiv \frac{\eta_{abk} + \eta^*_{ab\overline{k}}}{2}, \quad
    \eta^{''}_{abk} \equiv \frac{\eta_{abk} - \eta^*_{ab\overline{k}}}{2i},
\end{align}
and the cavity-loss superoperator
\begin{align}
    \mathcal{L}_\kappa \hat{O}
    =\kappa\left[
    \hat a_{\rm ph}\hat O\hat a_{\rm ph}^\dagger
    -\frac{1}{2}\left\{
    \hat a_{\rm ph}^\dagger\hat a_{\rm ph},\hat O
    \right\}\right].
\end{align}
The only Lindblad channel is cavity photon loss.  With this convention, $\kappa$ is the cavity spectral FWHM and the field amplitude decays at rate $\kappa/2$.

For the Drude--Lorentz spectral density $J(\omega) = 2\lambda\omega_f\omega/(\omega_f^2+\omega^2)$, we use the $[N/N]$ Pad\'{e} spectral decomposition (PSD)~\cite{Pade_2010, Pade_2011} of the Bose--Einstein distribution.  The bath TCF in Eq.~\ref{bare_bath_TCF} then becomes a sum of $K = 1 + N_p$ exponentials,
\begin{equation}
    C(t) = \eta_1 e^{-\omega_f t} + \sum_{k=2}^{K} \eta_k^{} e^{-\gamma_k^{} t},
\end{equation}
where the first term is the physical Drude pole at $\gamma_1 = \omega_f$.  The remaining $N_p$ Pad\'{e} terms converge with fewer terms than the Matsubara series.
The Pad\'{e} poles $\{\gamma_k\}$ and residues $\{\eta_k\}$ are determined by the $[N/N]$ rational approximant to the Bose function evaluated at complex argument $-i\omega_f/k_\mathrm{B}T$~\cite{Pade_2011}.

We use the same numerical parameters for the small-$N$ mixed HEOM--Lindblad calculations as in Ref.~\citenum{Ying_mn24}.
The Drude--Lorentz bath TCF uses the $[N/N]$ PSD scheme with $N_p = 4$ Pad\'{e} correction poles ($K = 5$ exponentials per Drude mode), hierarchy depth $L_\mathrm{max} = 20$, and dynamical filtering threshold $\varepsilon_\mathrm{HEOM} = 10^{-6}$.
The linear absorption spectrum is computed from the Fourier transform of the dipole--dipole time-correlation function,
\begin{equation}
    A(\omega) \propto \mathrm{Re}\int_0^\infty dt\, \langle\hat{\mu}(t)\hat{\mu}(0)\rangle\, e^{i\omega t},
\end{equation}
where $\langle\hat{\mu}(t)\hat{\mu}(0)\rangle = \mathrm{Tr}[\hat{\mu}(t)\hat{\mu}(0)\hat{\rho}_\mathrm{SS}]$ is evaluated with the HEOM steady-state reduced density matrix $\hat{\rho}_\mathrm{SS}$; see Sec.~V\,A of Ref.~\citenum{Ying_mn24} for details.

%%%%%%%%%%%%%%%%%%%%%%%%%%%%%%%%%%%%%%%%%%

\subsection{MF-HEOM and the Optical Response}\label{sec:SI_MF_HEOM}

The collective-limit benchmark uses the mean-field hierarchical equations of motion (MF-HEOM) introduced in Ref.~\citenum{Reichman_Nanoph2024} and the general HEOM construction summarized in Sec.~\ref{sec:I-A}.
The reduced cavity--molecule state is approximated by the product ansatz
\begin{equation}
\hat\rho(t)\simeq
\hat\rho_\mathrm{c}(t)\otimes
\prod_{j=1}^{N}\hat\rho_\mathrm{M}^{(j)}(t),
\label{eq:SI_MF_ansatz}
\end{equation}
with identical molecular states
$\hat\rho_\mathrm{M}^{(j)}(t)=\hat\rho_\mathrm{M}(t)$.
This approximation neglects connected intermolecular correlations but keeps their collective feedback through the cavity mean field
$\alpha(t)=\operatorname{Tr}[\hat\rho_\mathrm{c}(t)\hat a]$.

Each molecule and its local bath evolve under
\begin{equation}
\hat H_\mathrm{M}(\alpha,t)
=\hat H_\mathrm{S}(\alpha,t)+\hat h_\mathrm{B}+\hat H_\mathrm{SB},
\end{equation}
where
\begin{equation}
\hat H_\mathrm{S}(\alpha,t)
=\omega_0|e\rangle\!\langle e|
+g_\mathrm{c}\!\left[
\alpha^*(t)\hat\sigma^-+\alpha(t)\hat\sigma^+
\right]
-\hat\mu E(t).
\label{eq:SI_MF_HS}
\end{equation}
Here $\omega_0\equiv\omega_{00}+\lambda$ is the vertical transition energy used for the excited-state diagonal, $\hat\mu=\mu_0(\hat\sigma^-+\hat\sigma^+)$, and the last term sets the applied-field convention $\hat H_\mathrm{drive}=-\hat\mu E(t)$.
The bath Hamiltonian and the diagonal molecular coupling operator
$\hat Q=|e\rangle\!\langle e|$ are the same as in the main text.

Writing the bath correlation function as
$C_{ab}(t)=\sum_k\eta_{abk}e^{-\gamma_{ak}t}$, the molecular hierarchy is
\begin{align}
\dot\rho_{\mathbf n}^{(n)}(t)
={}&-\left[
i\mathcal L_\mathrm{S}(\alpha,t)
+\sum_{ak}n_{ak}\gamma_{ak}
\right]\rho_{\mathbf n}^{(n)}(t)
-i\sum_{ak}\mathcal Q_a^\times
\rho_{\mathbf n^+_{ak}}^{(n+1)}(t)\notag\\
&-i\sum_{abk}n_{ak}
\left(\eta'_{abk}\mathcal Q_b^\times
+i\eta''_{abk}\mathcal Q_b^\circ\right)
\rho_{\mathbf n^-_{ak}}^{(n-1)}(t).
\label{eq:SI_MF_HEOM}
\end{align}
The zeroth-tier auxiliary density operator is
$\rho_{\mathbf 0}^{(0)}(t)=\hat\rho_\mathrm{M}(t)$, and
\begin{align}
\mathcal L_\mathrm{S}(\alpha,t)\hat O
&\equiv[\hat H_\mathrm{S}(\alpha,t),\hat O],\quad
\mathcal Q_a^\times\hat O\equiv[\hat Q_a,\hat O],\quad
\mathcal Q_a^\circ\hat O\equiv\{\hat Q_a,\hat O\},\notag\\
\eta'_{abk}&\equiv
\frac{\eta_{abk}+\eta^*_{ab\bar k}}{2},\quad
\eta''_{abk}\equiv
\frac{\eta_{abk}-\eta^*_{ab\bar k}}{2i}.
\label{eq:SI_MF_defs}
\end{align}
For the present system–bath coupling channel,
$\hat Q_a=\hat Q=|e\rangle\!\langle e|$.

The molecular coherence that drives the cavity is
\begin{equation}
\sigma(t)=
\operatorname{Tr}\!\left[
\hat\rho_\mathrm{M}(t)\hat\sigma^-
\right],
\end{equation}
and the cavity mean field obeys
\begin{equation}
\dot\alpha(t)
=-i\left(\omega_\mathrm{c}-i\kappa/2\right)\alpha(t)
-ig_\mathrm{c}N\sigma(t).
\label{eq:SI_MF_cavity}
\end{equation}
Equations~\ref{eq:SI_MF_HEOM} and \ref{eq:SI_MF_cavity} are propagated together.

Initially every molecule is in $|g\rangle\!\langle g|$, all higher-tier auxiliary density operators vanish, and $\alpha(0)=0$.
A weak Gaussian pulse is applied,
\begin{equation}
E(t)=\frac{A\pi}{\sqrt{2\pi}\,\sigma}
\exp\!\left[-\frac{(t-t_0)^2}{2\sigma^2}\right]
\cos\!\left[\omega_\mathrm{p}(t-t_0)\right].
\label{eq:SI_MF_pulse}
\end{equation}
The response polarization is defined by
\begin{equation}
P(t)=\langle\hat\mu(t)\rangle-\mu_\mathrm{eq}.
\end{equation}
We use the discrete Fourier-transform convention
\begin{equation}
P(\omega)=\int dt\,P(t)e^{-i\omega t},
\qquad
E(\omega)=\int dt\,E(t)e^{-i\omega t},
\qquad
\chi(\omega)=\frac{P(\omega)}{E(\omega)}.
\label{eq:SI_MF_FT}
\end{equation}
Because
$\hat H_\mathrm{drive}=-\hat\mu E(t)$, the response kernel multiplying the
field is $+i\theta(t)\langle[\hat\mu(t),\hat\mu(0)]\rangle$, the negative of
the retarded susceptibility
$\chi_{\mu\mu}^R(t)=-i\theta(t)\langle[\hat\mu(t),\hat\mu(0)]\rangle$.
For Hermitian $\hat\mu$, changing from the $e^{+i\omega t}$ transform used
for $\chi_{\mu\mu}^R$ to the $e^{-i\omega t}$ transform in
Eq.~\ref{eq:SI_MF_FT} gives the following relation, where $(+)$ and $(-)$
label the sign in the Fourier exponential:
\begin{equation}
\chi_{\mathrm{MF}}^{(-)}(\omega)
=-[\chi_{\mu\mu}^{R,(+)}(\omega)]^*,
\qquad
-\operatorname{Im}\chi_{\mathrm{MF}}^{(-)}(\omega)
=-\operatorname{Im}\chi_{\mu\mu}^{R,(+)}(\omega).
\label{eq:SI_MF_observable}
\end{equation}
We compare both methods through the dipole spectral function
$-\operatorname{Im}\chi^R(\omega)$: MF-HEOM uses
$-\operatorname{Im}[P(\omega)/E(\omega)]$ and the rotating-wave calculation
uses $-\operatorname{Im}G_{BB}^R(\omega)$.  Absorbed power would include an
additional factor of $\omega$, which is not included here.

The pulse parameters are
$A=0.001\,\mathrm{a.u.}$,
$\omega_\mathrm{p}=2.0\,\mathrm{eV}$,
$\sigma=1\,\mathrm{fs}$, and
$t_0=5\,\mathrm{fs}$.
For the representative $\omega_\mathrm{c}=2.0\,\mathrm{eV}$ propagation,
the maximum excited-state population is $2.096\times10^{-6}$ and the maximum
trace error is $5.08\times10^{-12}$.  No pulse-amplitude series is available,
so these values establish weak excitation for this pulse but do not test
amplitude scaling.
The equations are propagated to $t_\mathrm{f}=4\,\mathrm{ps}$ with
$dt=0.01\,\mathrm{fs}$.
The Drude--Lorentz correlation function uses the $[N/N]$ Pad\'e decomposition with $N_p=4$ correction poles, hierarchy depth $L_\mathrm{max}=20$, and dynamical filtering threshold $\varepsilon_\mathrm{MF}=10^{-7}$.
All MF-HEOM spectra use $N=10^4$ to approximate the collective limit.
Section~\ref{sec:SI_collective_limit} gives the collective-limit comparison
and its normalization.

%%%%%%%%%%%%%%%%%%%%%%%%%%%%%%%%%%%%%%%%%%
\clearpage
\section{Green's-Function Derivation}\label{sec:SI_GF_details}

\subsection{Exact Zero- and One-Excitation Representation}
\label{sec:SI_exact_representation}

For the closed rotating-wave HTC Hamiltonian, the physical electronic--photonic
ground and one-excitation sectors are
\begin{equation}
\mathcal H_0=\operatorname{span}\{|G\rangle\},\qquad
\mathcal H_1=\operatorname{span}\{|C\rangle,|1\rangle,\ldots,|N\rangle\},
\label{eq:SI_excitation_spaces}
\end{equation}
where $|C\rangle=\hat a_{\rm ph}^\dagger|G\rangle$ and
$|n\rangle=\hat\sigma_n^+|G\rangle$.  These are electronic--photonic
sectors; each is tensored with the full vibrational Hilbert space.  In
particular, the physical reference state has thermal vibrations and need not
be the vibrational vacuum.

Introduce canonical bosonic operators $\hat a$ and $\hat d_n$ on an auxiliary
Fock space whose vacuum $|\mathrm{vac}\rangle$ represents $|G\rangle$.  The
subspace needed for ground-state linear absorption is
\begin{equation}
\mathcal F_{\rm exc}^{(0,1)}=\operatorname{span}\{|\mathrm{vac}\rangle,
\hat a^\dagger|\mathrm{vac}\rangle,\hat d_1^\dagger|\mathrm{vac}\rangle,
\ldots,\hat d_N^\dagger|\mathrm{vac}\rangle\},\qquad
\hat Q_{\rm exc}=\hat a^\dagger\hat a+\sum_n\hat d_n^\dagger\hat d_n
\in\{0,1\}.
\label{eq:SI_reduced_space}
\end{equation}
The mapping between these representations is
\begin{equation}
\mathcal V|G\rangle=|\mathrm{vac}\rangle,\qquad
\mathcal V|C\rangle=\hat a^\dagger|\mathrm{vac}\rangle,\qquad
\mathcal V|n\rangle=\hat d_n^\dagger|\mathrm{vac}\rangle.
\label{eq:SI_excitation_isometry}
\end{equation}
Here $\mathcal V$ acts as the identity on the vibrational Hilbert space.  The
bosonic operators are canonical on the full auxiliary Fock space and
reproduce the physical matrix elements within $Q_{\rm exc}=0,1$; the mapping
does not extend unchanged to higher excitation manifolds.

Writing
$\hat h_{\rm B}=\sum_{n\alpha}\omega_\alpha
\hat b_{\alpha n}^\dagger\hat b_{\alpha n}$ and
$\hat X_{\alpha n}=\hat b_{\alpha n}+\hat b_{\alpha n}^\dagger$, the matrix
elements of the original Hamiltonian in $\mathcal H_0\oplus\mathcal H_1$ are
exactly reproduced by
\begin{align}
\hat{\tilde H}={}&\hat h_{\rm B}
+\omega_{\rm c}\hat a^\dagger\hat a
+\omega_0\sum_n\hat d_n^\dagger\hat d_n
+g_{\rm c}\sum_n(\hat a^\dagger\hat d_n+\hat d_n^\dagger\hat a)\notag\\
&+\sum_{n\alpha}c_\alpha\hat d_n^\dagger\hat d_n\hat X_{\alpha n},
\qquad Q_{\rm exc}\in\{0,1\}.
\label{eq:SI_Hred}
\end{align}
Thus $\hat a$ and $\hat d_n$ are cavity- and molecular-excitation operators,
respectively.  Within the restricted sector their matrix elements equal those
of the corresponding physical transitions.  Every term in the closed
Hamiltonian conserves the excitation number,
\begin{equation}
[\hat{\tilde H},\hat Q_{\rm exc}]=0.
\label{eq:SI_Qexc_conservation}
\end{equation}
One optical creation takes the system from $Q_{\rm exc}=0$ to 1, the closed
Hamiltonian evolves it within that sector, and the final annihilation returns
it to $Q_{\rm exc}=0$.  This is why ground-state linear absorption requires
only the sectors in Eq.~\ref{eq:SI_reduced_space}.  Cavity leakage is added
separately through the retarded photon self-energy $-i\kappa/2$; the
conservation statement above refers to the closed HTC Hamiltonian.

The reference density used below is
\begin{equation}
\hat\rho_0=|\mathrm{vac}\rangle\langle\mathrm{vac}|
\otimes\hat\rho_{\rm B}^{\rm th}.
\label{eq:SI_reduced_reference}
\end{equation}
Together with the quadratic reference Hamiltonian
$\hat H_0\equiv\hat{\tilde H}_{\rm S}+\hat h_{\rm B}$, this is a Gaussian
reference, so Wick's theorem applies to the canonical cavity, molecular, and
phonon operators and their linear combinations.  A related
finite-Hilbert-space nonequilibrium Green's-function construction was
introduced for a single quantum-dot--cavity system in
Ref.~\citenum{Hornecker2017}, where an auxiliary-vacuum retarded Green's
function keeps the Kadanoff--Baym propagation linear in the physical state.

The sector mapping is exact; the approximation enters only when the
vibrational self-energy is evaluated with BA or SCBA.  The restricted
vacuum--one-excitation implementation cannot be used unchanged for nonlinear
response, initially excited populations, or Hamiltonians that connect to
higher excitation manifolds.

\subsection{Exciton--Phonon Vertices and the Fan Self-Energy}
\label{sec:GF_derivation}

The exact sector representation in Sec.~\ref{sec:SI_exact_representation}
defines the excitation vector
$\hat{\boldsymbol q}=(\hat a,\hat d_1,\ldots,\hat d_N)^T$.  Its
$(N+1)\times(N+1)$ retarded Green's-function matrix is
\begin{equation}
G_{ij}^R(t)=-i\theta(t)\operatorname{Tr}\!\left\{
\hat\rho_0[\hat q_i(t),\hat q_j^\dagger(0)]\right\},
\qquad i,j\in\{a,1,\ldots,N\},
\label{eq:SI_full_GF}
\end{equation}
and $\mathbf G_0^R$ denotes the same matrix with
$\hat{\tilde H}_{\rm SB}=0$.
For a uniform molecular probe, define the discrete Fourier combinations
\begin{equation}
\hat d_k=\frac{1}{\sqrt N}\sum_{n=1}^{N}
e^{-i2\pi kn/N}\hat d_n,
\qquad \hat d_0\equiv\hat B,
\qquad k=0,1,\ldots,N-1,
\label{eq:SI_collective_modes}
\end{equation}
whose inverse is
\begin{equation*}
\hat d_n=\frac{1}{\sqrt N}\sum_{k=0}^{N-1}
e^{i2\pi kn/N}\hat d_k.
\end{equation*}
Here $\hat B$ is bright and the $k\ne0$ modes are dark.
The integer $k$ labels an orthogonal collective combination; it is not a
physical crystal momentum, and no spatial periodicity of the molecules is
assumed.
In the site basis, the interaction vertex for phonon mode $(\alpha,n)$ is
\begin{equation}
\left(V^{\alpha n}\right)_{ij}
=c_\alpha\delta_{in}\delta_{jn},
\qquad
\hat{\tilde{H}}_\mathrm{SB}
=\sum_{\alpha n,ij}\hat q_i^\dagger
\left(V^{\alpha n}\right)_{ij}\hat q_j\hat X_{\alpha n},
\label{eq:SI_site_vertex}
\end{equation}
where the cavity row and column are zero.
For the phonons we use the same unitary convention,
\begin{equation*}
\hat b_{\alpha q}=\frac{1}{\sqrt N}\sum_{n=1}^{N}
e^{-i2\pi qn/N}\hat b_{\alpha n},\qquad
\hat b_{\alpha n}=\frac{1}{\sqrt N}\sum_{q=0}^{N-1}
e^{i2\pi qn/N}\hat b_{\alpha q}.
\end{equation*}
Writing $[r]_N$ for the residue of the integer $r$ modulo $N$, the inverse
transforms give
\begin{equation}
\hat{\tilde{H}}_\mathrm{SB}
=\frac{1}{\sqrt N}\sum_{\alpha}
\sum_{k,q=0}^{N-1}c_\alpha
\hat d_{[k+q]_N}^\dagger\hat d_k
(\hat b_{\alpha q}+\hat b_{\alpha,[-q]_N}^\dagger).
\label{eq:SI_collective_vertex}
\end{equation}
Thus
$V_{[k+q]_N,k}^{\alpha q}=c_\alpha/\sqrt N$, and the collective-mode label is
conserved modulo $N$ at every vertex.

\begin{figure}[htbp]
\centering
\input{Figures/SI/FigS1/FigS1.tex}
\caption{%
Diagrammatic content of the Fan/SCBA approximation.
Solid directed lines denote molecular-excitation propagators, blue dashed arcs denote the phonon propagator $D_0$, the short blue dashed stubs in (d) indicate external phonon legs, and black dots denote matter--phonon vertices.
(a)~SCBA uses the two-vertex Fan diagram with the self-consistent excitation propagator $\mathbf G$ and noninteracting phonon propagator $D_0$; setting $\mathbf G=\mathbf G_0$ gives the single-shot BA.
(b)~Dyson-expanding the self-consistent $\mathbf G$ in (a) generates the noncrossing rainbow diagrams included in SCBA.
(c)~Crossed phonon lines and vertex corrections are beyond SCBA.
(d)~The two-vertex phonon polarization is a closed excitation-density bubble.
Because the reduced excited-sector $G^<=0$ in the vacuum reference ensemble, $\Pi^{R,<,>}=0$ and $D^{R,<,>}=D_0^{R,<,>}$; the Hartree/tadpole vanishes as well.
}
\label{fig:SI_Fan_diagrams}
\end{figure}
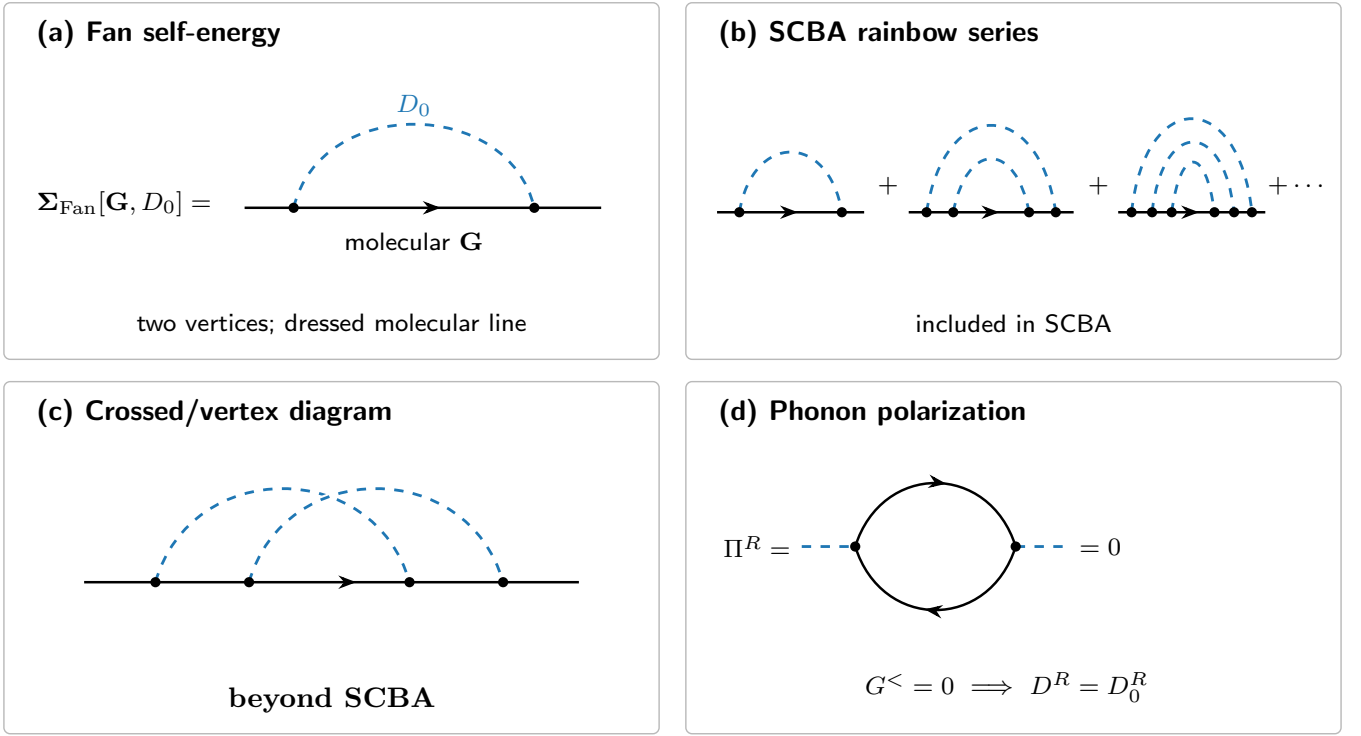

Writing Eq.~\ref{eq:SI_Hred} as $\hat{\tilde{H}}=\hat{\tilde H}_{\rm S}+\hat h_{\rm B}+\hat{\tilde H}_{\rm SB}$, with $\hat{\tilde H}_{\rm SB}$ defined by Eq.~\ref{eq:SI_site_vertex}, we use the Gaussian reference associated with Eq.~\ref{eq:SI_reduced_reference}.
Contracting two matter--phonon vertices gives the noncrossing Fan term.
SCBA uses this two-vertex diagram with a self-consistent internal molecular
propagator:
\begin{equation}
\boldsymbol\Sigma_\mathrm{Fan}(1,2)
=i\sum_{\alpha q}\boldsymbol V^{\alpha q}
\mathbf G(1,2)D_{0,\alpha q}(1,2)
\boldsymbol V^{\alpha,-q}.
\label{eq:SI_Fan_contour}
\end{equation}
The calculation uses the canonical bosonic excitation operators introduced in
Sec.~\ref{sec:SI_exact_representation}.
Figure~\ref{fig:SI_Fan_diagrams} shows which diagrams are included in this approximation.
Replacing the self-consistent $\mathbf G$ in Eq.~\ref{eq:SI_Fan_contour} with $\mathbf G_0$ gives the single-shot BA.
Expanding the dressed excitation propagator through its Dyson equation
generates the noncrossing rainbow series, but not crossed phonon lines or
vertex corrections.

For the reduced-vacuum reference state, the retarded convolution can be written
\begin{equation}
\boldsymbol\Sigma_\mathrm{Fan}^R(\omega)
=i\sum_{\alpha q}\int\frac{d\nu}{2\pi}\,
\boldsymbol V^{\alpha q}\mathbf G^R(\omega-\nu)
\boldsymbol V^{\alpha,-q}D_{0,\alpha}^>(\nu).
\label{eq:SI_Fan_convolution}
\end{equation}
Using the discrete harmonic spectrum gives
\begin{equation}
\widetilde\Sigma_{\mu\nu}^{R,(m+1)}(\omega)=\sum_{\alpha q,\rho\sigma}V_{\mu\rho}^{\alpha q}\Big[(\overline{n}(\omega_\alpha)+1)G_{\rho\sigma}^{R,(m)}(\omega-\omega_\alpha)+\overline{n}(\omega_\alpha)G_{\rho\sigma}^{R,(m)}(\omega+\omega_\alpha)\Big]V_{\sigma\nu}^{\alpha,-q}.
\label{eq:SI_SCBA_matrix}
\end{equation}
Here and below, $\overline{n}(\omega)=(e^{\beta\omega}-1)^{-1}$ denotes the Bose--Einstein distribution function.

Within the site-local SCBA, the molecular self-energy is identical on every
site.  In the ordered basis $(a,d_1,\ldots,d_N)$, the Dyson equation is
\begin{equation}
[\mathbf G^R(\omega)]^{-1}
=[\mathbf G_0^R(\omega)]^{-1}
-
\begin{pmatrix}
0&\mathbf 0\\
\mathbf 0&\Sigma^R(\omega)\mathbf 1_N
\end{pmatrix},
\label{eq:SI_full_Dyson}
\end{equation}
where $\mathbf 0$ denotes a zero block and $\mathbf 1_N$ is the identity in the molecular site subspace.  In the equivalent inhomogeneous Dyson equation, the right-hand side is the identity matrix: its $j$th column supplies the one-particle source $\hat q_j^\dagger|\mathrm{vac}\rangle$, so $G_{BB}^R$ is the response to a bright-state source.

The unitary transformation in Eq.~\ref{eq:SI_collective_modes} diagonalizes this uniform molecular self-energy.  Because only the bright mode couples to the cavity, the full inverse Green's function becomes the direct sum
\begin{equation}
[\mathbf G^R(\omega)]^{-1}
=[\mathbf G_{aB}^R(\omega)]^{-1}
\bigoplus_{k\ne0}[G_{D_k}^R(\omega)]^{-1}.
\label{eq:SI_direct_sum}
\end{equation}
The $N-1$ dark scalar blocks are degenerate,
$G_{D_k}^R\equiv G_D^R$, in this approximation; their scalar form and the
$2\times2$ cavity--bright block are given in the main text.  The internal
propagator in the self-energy includes the bright state and all $N-1$ dark
states,
\begin{equation}
\bar G^R(\omega)=
\frac{G_{BB}^R(\omega)+(N-1)G_D^R(\omega)}{N}.
\label{eq:SI_Gloc}
\end{equation}
The uniform matter self-energy therefore obeys
\begin{equation}
\widetilde\Sigma^{R,(m+1)}(\omega)=
\sum_\alpha c_\alpha^2\left[
(\overline{n}(\omega_\alpha)+1)\bar G^{R,(m)}(\omega-\omega_\alpha)
+\overline{n}(\omega_\alpha)\bar G^{R,(m)}(\omega+\omega_\alpha)\right].
\label{eq:SI_SCBA}
\end{equation}
For the continuous spectral density used in the calculations, the updated self-energy is
\begin{equation}
\widetilde\Sigma^{R,(m+1)}(\omega)=\frac{1}{\pi}\int_0^\infty d\nu\,J(\nu)
\left[(\overline{n}(\nu)+1)\bar G^{R,(m)}(\omega-\nu)
+\overline{n}(\nu)\bar G^{R,(m)}(\omega+\nu)\right].
\label{eq:SI_SCBA_integral}
\end{equation}

The single-shot Born approximation evaluates the same Fan self-energy kernel once with the noninteracting internal propagator
\begin{equation}
\bar G_0^R(z)=\frac{1}{N}\left[
G_{BB,0}^R(z)+\frac{N-1}{z-\omega_0+i0^+}
\right],
\label{eq:SI_Gbar0}
\end{equation}
where $G_{BB,0}^R$ is the bright component of the noninteracting cavity--bright block.
Explicitly,
\begin{equation}
\Sigma_\mathrm{BA}^R(\omega)=\frac{1}{\pi}\int_0^\infty d\nu\,J(\nu)\left[(\overline{n}(\nu)+1)\bar G_0^R(\omega-\nu)+\overline{n}(\nu)\bar G_0^R(\omega+\nu)\right].
\label{eq:SI_BA}
\end{equation}
This single-shot result initializes the SCBA iterations.

Section~\ref{sec:SI_phonon_GF} shows that the zero-excitation reference state
makes the phonon polarization and Hartree/tadpole vanish, so the Fan kernel
uses the bare phonon propagator.

The uniform site-diagonal scalar self-energy follows from this local two-vertex Fan/SCBA approximation.  Crossed or vertex diagrams can contain intersite propagators and generate off-site molecular elements.  For identical molecules with permutation-invariant bath statistics, the general form
\begin{equation}
\Sigma_{nm}^R=\Sigma_d^R\delta_{nm}+\Sigma_o^R(1-\delta_{nm})
\label{eq:SI_permutation_self_energy}
\end{equation}
still preserves the irreducible sectors: $\Sigma_{\mathrm{br}}^R=\Sigma_d^R+(N-1)\Sigma_o^R$ is the bright value and $\Sigma_{\mathrm{dk}}^R=\Sigma_d^R-\Sigma_o^R$ is $(N-1)$-fold degenerate.  Higher-order diagrams can therefore split the bright and dark self-energies without necessarily eliminating the symmetry reduction.

\subsection{Phonon Green's Functions in the Zero-Excitation Reference State}
\label{sec:SI_phonon_GF}

For the phonon displacement operator
\begin{equation}
\hat X_{\alpha n}
=\hat b_{\alpha n}+\hat b_{\alpha n}^\dagger,
\label{eq:SI_phonon_displacement}
\end{equation}
the retarded phonon Green's function is
\begin{equation}
D_{\alpha n,\beta m}^R(t)
=-i\theta(t)\operatorname{Tr}\!\left\{
\hat\rho_0[\hat X_{\alpha n}(t),\hat X_{\beta m}(0)]
\right\}.
\label{eq:SI_phonon_GF}
\end{equation}
The retarded and advanced functions of one uncoupled mode are
\begin{align}
D_{0,\alpha}^R(\nu)
&=\frac{1}{\nu-\omega_\alpha+i0^+}
-\frac{1}{\nu+\omega_\alpha+i0^+},\\
D_{0,\alpha}^A(\nu)&=\left[D_{0,\alpha}^R(\nu)\right]^*.
\label{eq:SI_D0RA}
\end{align}
They contain no Bose occupation and are temperature independent.
The lesser and greater components are
\begin{subequations}\label{eq:SI_D0lessgreater}
\begin{align}
D_{0,\alpha}^<(\nu)&=-2\pi i\left[
\overline{n}(\omega_\alpha)\delta(\nu-\omega_\alpha)
+(\overline{n}(\omega_\alpha)+1)\delta(\nu+\omega_\alpha)\right],\\
D_{0,\alpha}^>(\nu)&=-2\pi i\left[
(\overline{n}(\omega_\alpha)+1)\delta(\nu-\omega_\alpha)
+\overline{n}(\omega_\alpha)\delta(\nu+\omega_\alpha)\right].
\end{align}
\end{subequations}
Equivalently, the unsymmetrized bath correlation appearing in HEOM is
\begin{equation}
C_\alpha(t)=c_\alpha^2\langle\hat X_\alpha(t)\hat X_\alpha(0)\rangle
=c_\alpha^2\left[(\overline{n}(\omega_\alpha)+1)e^{-i\omega_\alpha t}
+\overline{n}(\omega_\alpha)e^{i\omega_\alpha t}\right].
\label{eq:SI_bath_correlation}
\end{equation}

The retarded phonon Dyson equation is
\begin{equation}
[D^R(\nu)]^{-1}
=[D_0^R(\nu)]^{-1}-\Pi^R(\nu),
\label{eq:SI_D_Dyson}
\end{equation}
where $\Pi^R$ is the phonon polarization.  At the two-vertex bubble level
corresponding to the Fan construction, its contour contribution is
\begin{equation}
\Pi_{\alpha n,\beta m}(1,2)
=i\,c_\alpha c_\beta
G_{nm}(1,2)G_{mn}(2,1),
\label{eq:SI_Pi_contour}
\end{equation}
The product of two single-particle propagators in
Eq.~\ref{eq:SI_Pi_contour} is the bubble approximation, not a general exact
polarization formula for an occupied interacting system.
The associated operator response is the connected excitation-density
correlator
\begin{equation}
\Pi_{\alpha n,\beta m}^R(t)
=-i\theta(t)c_\alpha c_\beta
\left\langle\left[
\hat d_n^\dagger(t)\hat d_n(t),
\hat d_m^\dagger(0)\hat d_m(0)
\right]\right\rangle_\mathrm{conn}.
\label{eq:SI_Pi_operator}
\end{equation}
In the absorption reference state,
\begin{equation}
\hat d_n^\dagger\hat d_n|\mathrm{vac}\rangle=0,\qquad
\langle\hat d_n^\dagger\hat d_n\rangle=0,\qquad G_{nn}^<=0.
\label{eq:SI_vacuum}
\end{equation}
Every density correlator in Eq.~\ref{eq:SI_Pi_operator} is therefore zero.
Hence
\begin{equation}
\Pi^{R,<,>}=0,\qquad
D^{R,<,>}=D_0^{R,<,>}.
\label{eq:SI_Dbare}
\end{equation}
The Hartree/tadpole is also proportional to $\langle\hat d_n^\dagger\hat d_n\rangle$ and vanishes.
Thermal phonons still enter the Fan kernel through the Bose emission and absorption factors $\overline{n}(\nu)+1$ and $\overline{n}(\nu)$; $\overline{n}(\nu)>0$ does not imply a reduced excited-state population.

\subsection{Equivalent Polariton-Basis Representation}
\label{sec:SI_polariton_basis}

The numerical calculation is performed in the cavity--bright block with the $(\hat a,\hat B)$ basis.  
An equivalent representation can be achieved by defining $\hat{\boldsymbol q}_{aB}\equiv(\hat a,\hat B)^T,\,\hat{\boldsymbol q}_{\mathrm{pol}}\equiv(\hat P_+,\hat P_-)^T$, and 
\begin{equation}
\begin{aligned}
\hat{\boldsymbol q}_{\mathrm{pol}}
&=U\hat{\boldsymbol q}_{aB},\qquad
U=
\begin{pmatrix}
\sin\Theta&\cos\Theta\\
\cos\Theta&-\sin\Theta
\end{pmatrix},
\end{aligned}
\label{eq:SI_polariton_transform}
\end{equation}
where $U^{-1}=U^T$.
With $\Delta=\omega_\mathrm{c}-\omega_0$ and
$\Omega_\mathrm{R}=(\Delta^2+4Ng_\mathrm{c}^2)^{1/2}$, the angle $\Theta$ is determined by
\begin{equation}
\cos(2\Theta)=-\frac{\Delta}{\Omega_\mathrm{R}},\qquad
\sin(2\Theta)=\frac{2\sqrt N g_\mathrm{c}}{\Omega_\mathrm{R}},\qquad
0\leq\Theta\leq\frac{\pi}{2},
\label{eq:SI_hopfield_branch}
\end{equation}
and $\omega_\pm=(\omega_0+\omega_\mathrm{c}\pm\Omega_\mathrm{R})/2$.
This convention gives $\Theta\to0$ for an exciton-like UP at large negative detuning, $\Theta\to\pi/2$ for a cavity-like UP at large positive detuning, and $\Theta=\pi/4$ at resonance.

The cavity- and matter-amplitude vectors in the polariton basis are
\begin{equation}
\mathbf v=(\sin\Theta,\cos\Theta)^T,\qquad
\mathbf u=(\cos\Theta,-\sin\Theta)^T,
\qquad
\hat a=\mathbf v^T\hat{\boldsymbol q}_{\mathrm{pol}},\quad
\hat B=\mathbf u^T\hat{\boldsymbol q}_{\mathrm{pol}}.
\label{eq:SI_amplitude_vectors}
\end{equation}
The interacting and noninteracting Green's functions transform as
\begin{subequations}
\label{eq:SI_Gpol_transform}
\begin{align}
\mathbf G_{\mathrm{pol}}^R(\omega)
&=U\mathbf G_{aB}^R(\omega)U^T,\\
\mathbf G_{\mathrm{pol},0}^R(\omega)
&=U\mathbf G_{aB,0}^R(\omega)U^T.
\end{align}
\end{subequations}
Equivalently, the noninteracting inverse propagator is
\begin{equation}
\left[\mathbf G_{\mathrm{pol},0}^R(\omega)\right]^{-1}=U\left[\mathbf G_{aB,0}^R(\omega)\right]^{-1}U^T=\begin{pmatrix}\omega-\omega_++i0^+&0\\0&\omega-\omega_-+i0^+\end{pmatrix}+\frac{i\kappa}{2}\mathbf v\mathbf v^T.
\label{eq:SI_Gpol0}
\end{equation}
Thus the noninteracting polariton propagator is diagonal (for $\kappa=0$), whereas finite cavity loss generates a generally non-diagonal projector $\mathbf v\mathbf v^T$.

The polariton Dyson equation is
\begin{equation}
\left[\mathbf G_{\mathrm{pol}}^R(\omega)\right]^{-1}
=\left[\mathbf G_{\mathrm{pol},0}^R(\omega)\right]^{-1}
-\boldsymbol\Sigma_{\mathrm{pol}}^R(\omega),
\label{eq:SI_Gpol_Dyson}
\end{equation}
where projection of the uniform matter self-energy gives
\begin{equation}
\boldsymbol\Sigma_{\mathrm{pol}}^R(\omega)
=\Sigma^R(\omega)\mathbf u\mathbf u^T
=\Sigma^R(\omega)
\begin{pmatrix}
\cos^2\Theta&-\sin\Theta\cos\Theta\\
-\sin\Theta\cos\Theta&\sin^2\Theta
\end{pmatrix}.
\label{eq:Sigma_SI}
\end{equation}
The Hamiltonian part of Eq.~\ref{eq:SI_Gpol0} is diagonal, but the matter Fan self-energy is generally non-diagonal.  This is equivalent to the cavity--bright representation.

Finally, Eq.~\ref{eq:SI_amplitude_vectors} gives the optical-response identity
\begin{equation}
G_{BB}^R(\omega)
=\mathbf u^T\mathbf G_{\mathrm{pol}}^R(\omega)\mathbf u,\qquad
A(\omega)\propto-\operatorname{Im}G_{BB}^R(\omega)
=-\operatorname{Im}\!\left[
\mathbf u^T\mathbf G_{\mathrm{pol}}^R(\omega)\mathbf u
\right].
\label{eq:SI_absorption_identity}
\end{equation}
The first expression is used throughout the numerical calculations, while the second demonstrates basis equivalence.

\subsection{Relation to the Standard Pseudoparticle Construction}
\label{sec:SI_PP_reduction}

For comparison with the standard pseudoparticle (PP) formulation, introduce
the unphysical PP vacuum $|\mathrm{PP}\rangle_{\rm vac}$ and one operator
$\hat p_A$ for each retained many-body state
$A\in\{G,C,1,\ldots,N\}$.  Physical states and Hubbard operators
$\hat{\mathcal X}_{AB}=|A\rangle\langle B|$ are represented
as~\cite{White_2012,White_2014,Galperin_2015}
\begin{equation}
|A\rangle\longleftrightarrow
\hat p_A^\dagger|\mathrm{PP}\rangle_{\rm vac},\qquad
\hat{\mathcal X}_{AB}\longleftrightarrow\hat p_A^\dagger\hat p_B,
\qquad
\hat Q_{\rm PP}\equiv
\hat p_G^\dagger\hat p_G+\hat p_C^\dagger\hat p_C
+\sum_n\hat p_n^\dagger\hat p_n=1.
\label{eq:SI_standard_PP_map}
\end{equation}
The state and Hubbard-operator correspondences are identities only after
restriction to the physical $Q_{\rm PP}=1$ sector.  With an arbitrary
ground-state energy $E_G$, the projected HTC Hamiltonian is
\begin{align}
\hat H_{\rm PP}={}&\hat h_{\rm B}+E_G\hat Q_{\rm PP}
+\omega_{\rm c}\hat p_C^\dagger\hat p_C
+\omega_0\sum_n\hat p_n^\dagger\hat p_n\notag\\
&+g_{\rm c}\sum_n
\left(\hat p_C^\dagger\hat p_n+\hat p_n^\dagger\hat p_C\right)
+\sum_{n\alpha}c_\alpha\hat p_n^\dagger\hat p_n\hat X_{\alpha n}.
\label{eq:SI_standard_PP_H}
\end{align}
Because $E_G\hat Q_{\rm PP}=E_G$ in the physical sector, this term is a
removable constant; we set $E_G=0$.

The PP vacuum is unphysical, whereas the excitation vacuum represents the
physical electronic--photonic ground state (with its thermal vibrational
density).  The occupied ground-state PP maps to that excitation vacuum by
\begin{equation}
\mathcal U\hat p_G^\dagger|\mathrm{PP}\rangle_{\rm vac}
=|\mathrm{vac}\rangle,\qquad
\mathcal U\hat p_C^\dagger|\mathrm{PP}\rangle_{\rm vac}
=\hat a^\dagger|\mathrm{vac}\rangle,\qquad
\mathcal U\hat p_n^\dagger|\mathrm{PP}\rangle_{\rm vac}
=\hat d_n^\dagger|\mathrm{vac}\rangle.
\label{eq:SI_PP_isometry}
\end{equation}
Both restricted cavity--matter spaces have dimension $N+2$, so $\mathcal U$
is unitary between them and acts as the identity on the vibrational bath.
Within these restricted spaces, the relevant bilinears obey
\begin{align}
\hat p_G^\dagger\hat p_G&\longleftrightarrow1-\hat Q_{\rm exc},&
\hat p_C^\dagger\hat p_G&\longleftrightarrow\hat a^\dagger,&
\hat p_n^\dagger\hat p_G&\longleftrightarrow\hat d_n^\dagger,\notag\\
\hat p_C^\dagger\hat p_C&\longleftrightarrow\hat a^\dagger\hat a,&
\hat p_C^\dagger\hat p_n&\longleftrightarrow\hat a^\dagger\hat d_n,&
\hat p_n^\dagger\hat p_m&\longleftrightarrow\hat d_n^\dagger\hat d_m.
\label{eq:SI_PP_bilinears}
\end{align}
Applying this isometry to Eq.~\ref{eq:SI_standard_PP_H} gives
Eq.~\ref{eq:SI_Hred}.  In particular,
$\hat p_n^\dagger\hat p_G\longleftrightarrow\hat d_n^\dagger$; the operators
used in Secs.~\ref{sec:GF_derivation}--\ref{sec:SI_polariton_basis} are the
resulting canonical excitation operators.

All retained cavity--matter states have the same electron number and hence
the same fermion-number parity.  They therefore have the same PP statistics.
Bosonic and fermionic PP realizations give the same
restricted matrix elements and open-line vacuum propagator; the bosonic
choice is convenient here.  For nonlinear response, finite excited
populations, or multi-excitation dynamics, the reduced construction of
Sec.~\ref{sec:SI_exact_representation} cannot be used unchanged.  A standard
PP treatment is one possible starting point, provided the physical
$Q_{\rm PP}=1$ constraint, occupied-state Keldysh components, and all required
many-body states are retained.

%%%%%%%%%%%%%%%%%%%%%%%%%%%%%%%%%%%%%%%%%%
\clearpage
\section{Benchmarks for the Site-Local SCBA}
\label{sec:SI_validity_benchmarks}

The isolated-molecule problem isolates errors in the multiphonon weights of a displaced oscillator; the finite-ensemble extension tests the same local approximation in the presence of cavity-induced bright and dark sectors.
The dimer then shows how direct intersite coupling adds bright--dark conversion and modulation when the electronic and vibrational splittings are commensurate.
% Calling this modulation secondary specifies the logical validity test---small local $S$ first, then detuning from allowed vibronic conversions---and does not require a tuned finite-$J$ enhancement to be numerically smaller than the baseline.

\subsection{Isolated-Molecule Benchmarks}
\label{sec:SI_single_site_validity}

\paragraph{Exact independent-boson reference.}
\label{sec:single_site_exact}

We first benchmark a single two-level molecule ($N=1$, $g_\mathrm{c}=0$) with diagonal system--bath coupling.
For a harmonic Gaussian bath, the independent-boson cumulant gives the exact molecular response
\begin{equation}
A(\omega)\propto\operatorname{Re}\int_0^\infty dt\,e^{i\omega t}C(t),
\label{eq:SI_cumulant_A}
\end{equation}
which can be derived by a full Lang--Firsov transformation.
Here $\hat d\equiv\hat\sigma^-$ is the physical two-level transition
operator, so $\hat d^\dagger\hat d=|e\rangle\langle e|$ is a projector in
the physical molecular Hilbert space.  The Frenkel Hamiltonian is
\begin{equation}
\hat H_\mathrm{F}
=\omega_0\hat d^\dagger\hat d
+\sum_\alpha\omega_\alpha\hat b_\alpha^\dagger\hat b_\alpha
+\hat d^\dagger\hat d\sum_\alpha c_\alpha
(\hat b_\alpha+\hat b_\alpha^\dagger).
\label{eq:SI_cumulant_H}
\end{equation}
We use the energy and spectral-density conventions
\begin{equation}
\omega_0=\omega_{00}+\lambda,\qquad
J(\nu)=\pi\sum_\alpha c_\alpha^2\delta(\nu-\omega_\alpha),
\qquad
\lambda=\sum_\alpha\frac{c_\alpha^2}{\omega_\alpha}
=\frac{1}{\pi}\int_0^\infty d\nu\,\frac{J(\nu)}{\nu}.
\label{eq:SI_cumulant_conventions}
\end{equation}
Using the physical two-level identity
$(\hat d^\dagger\hat d)^2=\hat d^\dagger\hat d$ (not an identity for an
unrestricted canonical boson), the full-polaron transformation is
\begin{equation}
\hat U=\exp\!\left[
\hat d^\dagger\hat d\sum_\alpha\frac{c_\alpha}{\omega_\alpha}
(\hat b_\alpha^\dagger-\hat b_\alpha)
\right].
\label{eq:SI_polaron_U}
\end{equation}
It removes the linear coupling exactly, while dressing the physical transition dipole $\hat\mu=\mu_0(\hat d+\hat d^\dagger)$:
\begin{equation}
\hat U\hat H_\mathrm{F}\hat U^\dagger
=\omega_{00}\hat d^\dagger\hat d
+\sum_\alpha\omega_\alpha\hat b_\alpha^\dagger\hat b_\alpha,
\label{eq:SI_polaron_H}
\end{equation}
\begin{equation}
\hat U\hat\mu\hat U^\dagger
=\mu_0\!\left[
\hat d\exp\!\left(-\sum_\alpha\frac{c_\alpha}{\omega_\alpha}
(\hat b_\alpha^\dagger-\hat b_\alpha)\right)
+\hat d^\dagger\exp\!\left(\sum_\alpha\frac{c_\alpha}{\omega_\alpha}
(\hat b_\alpha^\dagger-\hat b_\alpha)\right)
\right].
\label{eq:SI_polaron_mu}
\end{equation}
The absorption reference state
\begin{equation}
\hat\rho_0=|g\rangle\langle g|\otimes\hat\rho_\mathrm{B}^\mathrm{th}
\label{eq:SI_cumulant_rho0}
\end{equation}
is unchanged by $\hat U$ because $\hat d^\dagger\hat d|g\rangle=0$.
The normalized absorption part of the dipole correlation is therefore
\begin{equation}
\begin{aligned}
C(t)&\equiv\mu_0^{-2}
\langle\hat\mu(t)\hat\mu(0)\rangle_{\hat\rho_0}=e^{-i\omega_{00}t}
\left\langle
\exp\!\left[-\sum_\alpha\frac{c_\alpha}{\omega_\alpha}
(\hat b_\alpha^\dagger e^{i\omega_\alpha t}
-\hat b_\alpha e^{-i\omega_\alpha t})\right]
\exp\!\left[\sum_\alpha\frac{c_\alpha}{\omega_\alpha}
(\hat b_\alpha^\dagger-\hat b_\alpha)\right]
\right\rangle_\mathrm{B}.
\end{aligned}
\label{eq:SI_cumulant_displacement}
\end{equation}
The thermal bath is Gaussian, so this displacement average is exact.  Evaluating it and using Eq.~\ref{eq:SI_cumulant_conventions} gives
\begin{equation}
C(t)=\exp[-i\omega_0t-g(t)],
\label{eq:SI_cumulant_C}
\end{equation}
where
\begin{equation}
g(t)=\frac{1}{\pi}\int_0^\infty d\nu\,\frac{J(\nu)}{\nu^2}
\left[
(1-\cos\nu t)\coth\!\left(\frac{\beta\nu}{2}\right)
+i(\sin\nu t-\nu t)
\right].
\label{eq:SI_cumulant_g}
\end{equation}
Thus Eq.~\ref{eq:SI_cumulant_C} is exactly the correlation function of the polaron-dressed dipole.
% For the Drude--Lorentz benchmark in Sec.~\ref{sec:single_site_drude}, the same response is evaluated independently with the converged HEOM protocol of Sec.~\ref{sec:I-A}; the agreement between HEOM and the cumulant result checks both exact references numerically.

For one boson of frequency $\nu_0$ at $T=0$,
\begin{equation}
J_1(\omega')=\pi S\nu_0^2\delta(\omega'-\nu_0),
\qquad g=\nu_0\sqrt S,
\qquad \lambda=\frac{g^2}{\nu_0}=S\nu_0.
\label{eq:SI_single_boson_J}
\end{equation}
Equations~\ref{eq:SI_cumulant_C}--\ref{eq:SI_cumulant_g} then reduce analytically to
\begin{equation}
C_1(t)=e^{-i\omega_{00}t}e^{-S(1-e^{-i\nu_0 t})}
=e^{-S}\sum_{n=0}^{\infty}\frac{S^n}{n!}e^{-i(\omega_{00}+n\nu_0)t},
\label{eq:SI_single_boson_C}
\end{equation}
and, with Lorentzian half-width $\eta$,
\begin{equation}
A_1(\omega)\propto\sum_{n=0}^{\infty}e^{-S}\frac{S^n}{n!}
\frac{\eta}{[\omega-(\omega_{00}+n\nu_0)]^2+\eta^2}.
\label{eq:SI_single_boson_A}
\end{equation}
This is the Franck--Condon progression generated by the phonon-displacement correlation in a Lang--Firsov (polaron) transformation~\cite{Nitzan}: expanding the polaron dressing factor gives the same Poisson weights.
% More generally, the derivation in Sec.~\ref{sec:single_site_exact} shows that the full-polaron and cumulant solutions coincide for any harmonic Gaussian bath when $g_\mathrm{c}=0$; with a cavity, the exact reference is instead obtained by diagonalizing the coupled photon--vibronic Hamiltonian.

\paragraph{Single-mode vibration.}
For an isolated mode at $T=0$, let $z=\omega+i\eta$ and
$g=\nu_0\sqrt S$.  Iterating the site-local SCBA equation and resolving
the exact excited-state oscillator give, respectively,
\begin{align}
G_{\mathrm{SCBA}}^R(z)
&=
\cfrac{1}{z-\omega_0-
 \cfrac{g^2}{z-\omega_0-\nu_0-
  \cfrac{g^2}{z-\omega_0-2\nu_0-
   \cfrac{g^2}{\ddots}}}},
\label{eq:SI_single_mode_SCBA_cf}\\[4pt]
G_{\mathrm{ex}}^R(z)
&=
\cfrac{1}{z-\omega_0-
 \cfrac{g^2}{z-\omega_0-\nu_0-
  \cfrac{2g^2}{z-\omega_0-2\nu_0-
   \cfrac{3g^2}{\ddots}}}}.
\label{eq:SI_single_mode_exact_cf}
\end{align}
The exact oscillator matrix elements $g\sqrt{n+1}$ produce successive
numerators $(n+1)g^2$, whereas SCBA recursively reinserts the same two-vertex
Fan kernel and repeats $g^2$ at every rung.  The expressions agree through
$O(g^2)$ and first differ at $O(g^4)$, so self-consistency alone cannot
reconstruct the exact multiphonon weights.  For a continuous bath, accuracy
also depends on its spectrum, damping, temperature, and the system energy
gaps.  The finite-$J$ scan in Sec.~\ref{sec:SI_dimer_validity} uses
$\eta/\nu_0=0.005$ and its own $J=0$ reference.

\begin{figure}[htbp]
    \centering
    \includegraphics[width=1.0\linewidth]{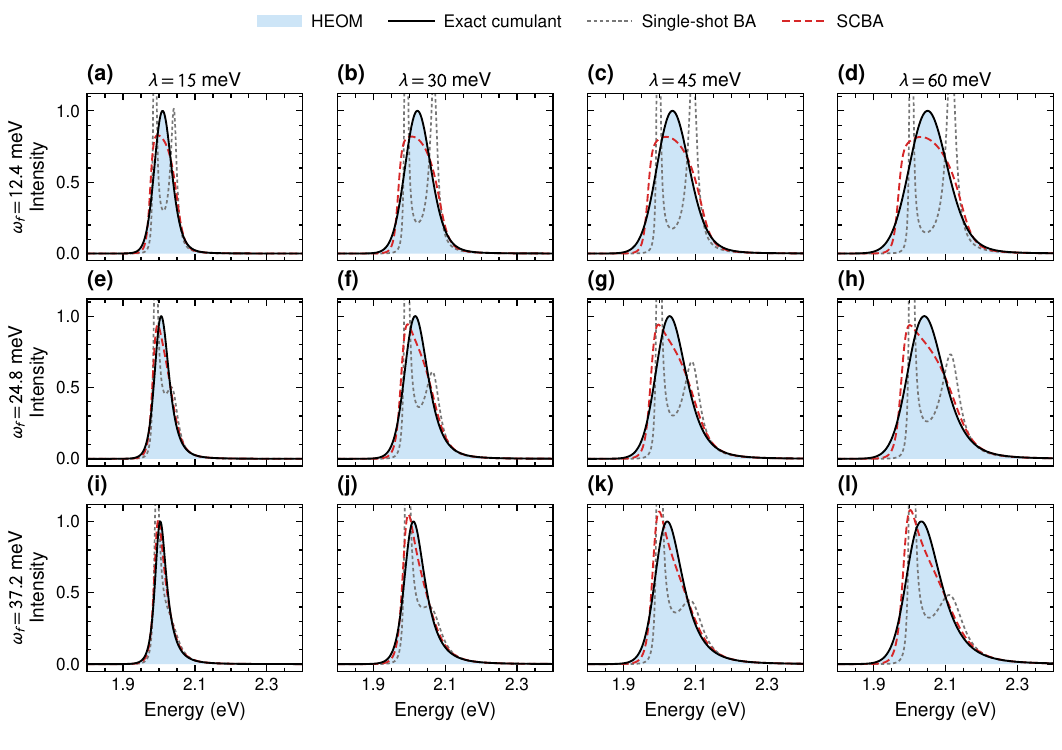}
    \caption{%
    Frenkel-exciton benchmark ($N=1$, $g_\mathrm{c}=0$, $\omega_{00}=2\,\mathrm{eV}$, $T=300\,\mathrm{K}$) for the Drude--Lorentz spectral density $J(\nu)=2\lambda\omega_f\nu/(\nu^2+\omega_f^2)$.
    HEOM is shown as light-blue shading, the exact cumulant result as a black solid curve, single-shot BA as a gray short-dashed curve, and SCBA as a red dashed curve.
    HEOM, BA, and SCBA are each matched to the exact-cumulant integrated area;
    the exact-cumulant peak sets their common vertical scale.
    Columns correspond to reorganization energies $\lambda = 15, 30, 45, 60\,\mathrm{meV}$ (left to right); rows correspond to bath cutoff frequencies $\omega_f = 12.4, 24.8, 37.2\,\mathrm{meV}$ (top to bottom).
    }
    \label{fig:single_site_drude_benchmark}
\end{figure}

\paragraph{Drude--Lorentz bath.}
\label{sec:single_site_drude}
Figure~\ref{fig:single_site_drude_benchmark} uses
$J(\nu)=2\lambda\omega_f\nu/(\nu^2+\omega_f^2)$, where $\lambda$ is the
reorganization energy and $\omega_f^{-1}$ is the bath correlation time.

\FloatBarrier
At $T=300\,\mathrm{K}$, SCBA reproduces the central peak and broadening most
closely at smaller $\lambda$; peak-position and asymmetry errors grow with
coupling.  BA remains narrower because its internal $G_0^R$ has no
phonon-induced lifetime, whereas SCBA feeds the dressed linewidth back into
Eq.~\ref{eq:SI_SCBA_integral}.

\begin{figure}[htbp]
    \centering
    \includegraphics[width=1.0\linewidth]{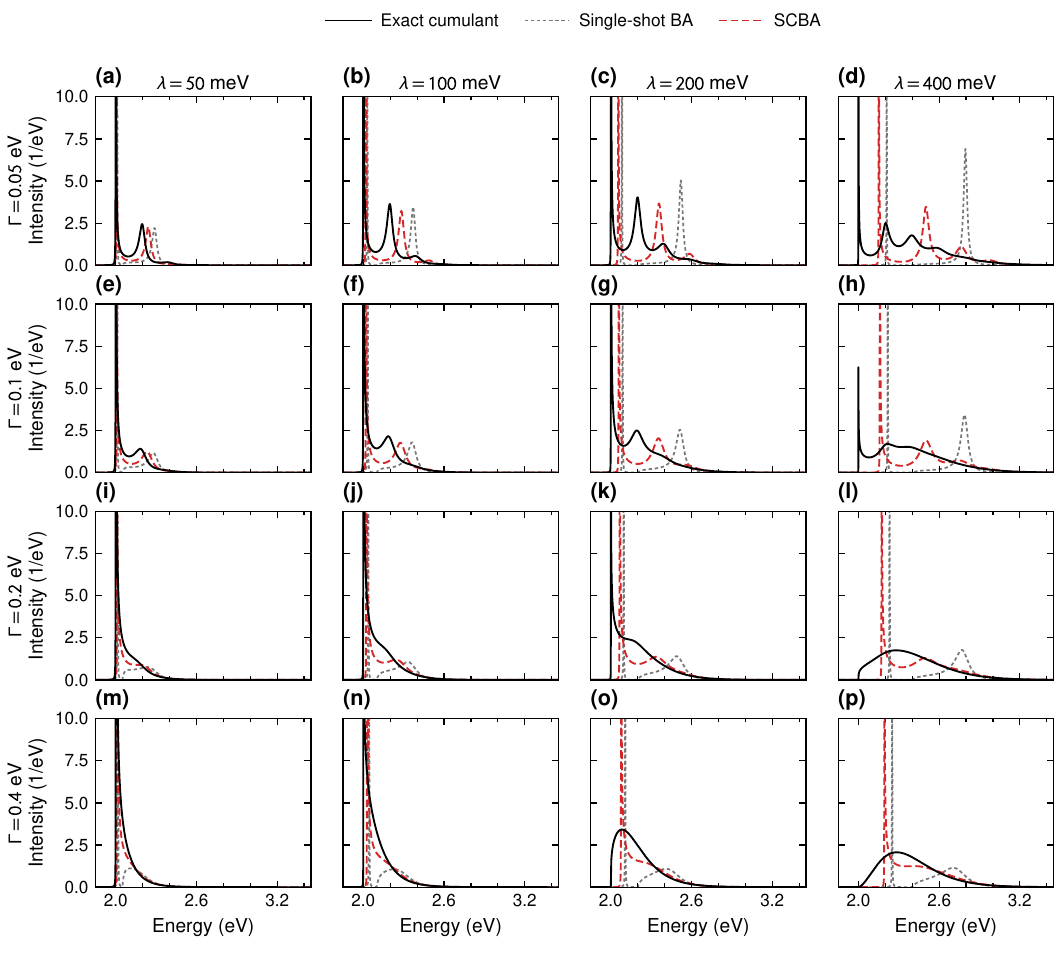}
    \caption{%
    Zero-temperature Frenkel-exciton benchmark ($N=1$, $g_\mathrm{c}=0$, $\omega_{00}=2.0\,\mathrm{eV}$) for the Brownian spectral density used here, with $\omega_s=0.2\,\mathrm{eV}$ and retarded half-width $\eta=0.1\,\mathrm{meV}$.
    Columns correspond to $\lambda=50$, $100$, $200$, and $400\,\mathrm{meV}$ (left to right), and rows correspond to $\Gamma=0.05$, $0.1$, $0.2$, and $0.4\,\mathrm{eV}$ (top to bottom).
    The exact independent-boson cumulant is black solid, single-shot BA is gray short-dashed, and SCBA is red dashed.
    Each spectrum has unit area over $1.6$--$3.8\,\mathrm{eV}$.
    A common linear vertical range of $0$--$10\,\mathrm{eV}^{-1}$ is used
    to expose the broad backgrounds; narrower poles extend above this displayed
    range.  Weak damping resolves vibronic structure, whereas increasing
    $\Gamma$ produces a broader continuum.
    }
    \label{fig:brownian_benchmark}
\end{figure}

\paragraph{Brownian bath.}
\label{sec:single_site_brownian}
Figure~\ref{fig:brownian_benchmark} tests
$J(\nu)=2\lambda\Gamma\omega_s^2\nu/
[(\nu^2-\omega_s^2)^2+\Gamma^2\nu^2]$.
Here $\omega_s=0.2\,\mathrm{eV}$, $T=0$, and
$\Gamma=0.05$--$0.4\,\mathrm{eV}$ spans the underdamped to critically
damped regimes.

\FloatBarrier
At weak damping, BA misses exact multiphonon weights and SCBA better follows
the overall line shape.  Increasing $\Gamma$ spreads the bath weight into a
continuum; method differences remain largest at strong coupling.

\subsection{Discrete-Mode Cavity Ensembles}
\label{sec:single_site_discrete}
\begin{figure}[!htbp]
    \centering
    \includegraphics[width=0.95\linewidth]{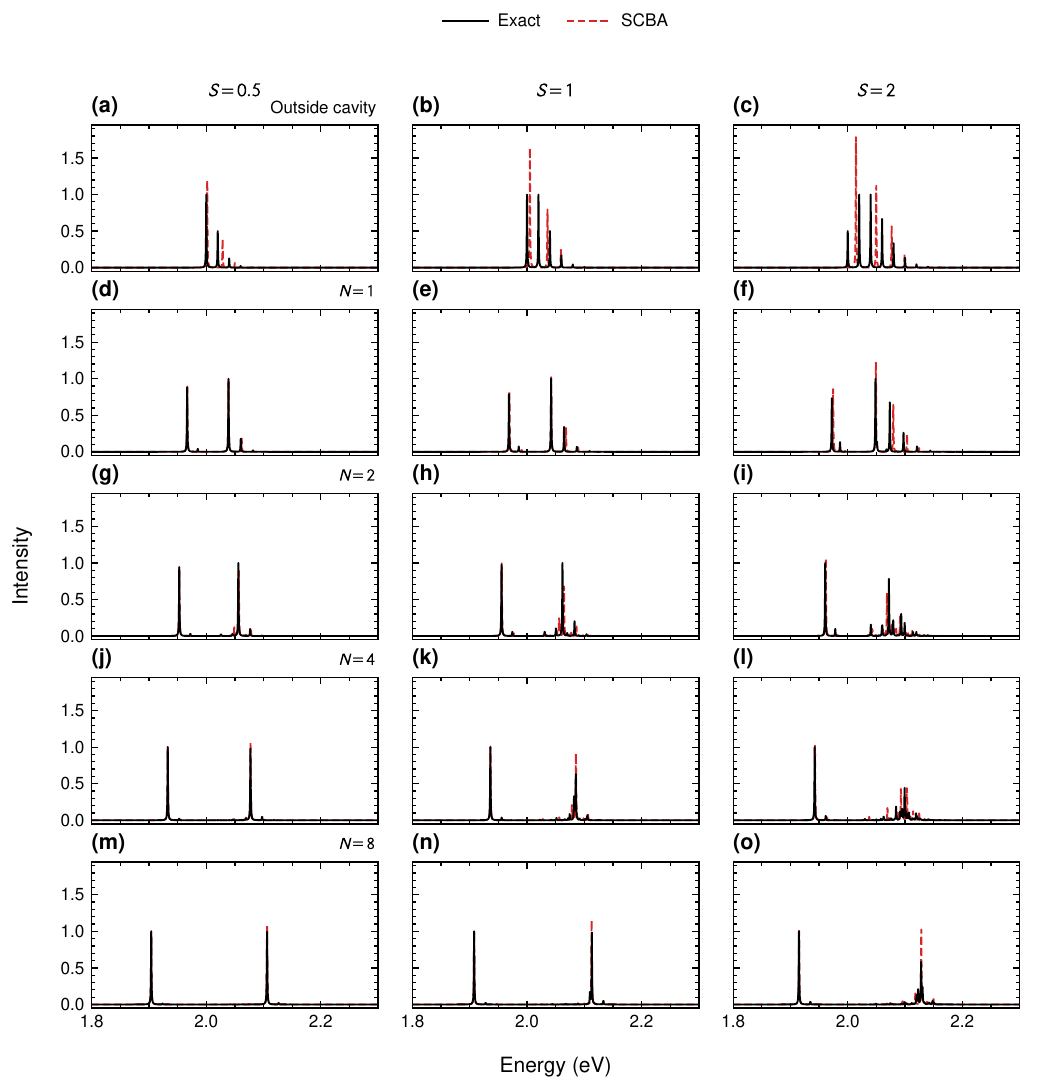}
    \caption{%
    Discrete-mode benchmark at 0--0 resonance,
    $\omega_\mathrm{c}=\omega_{00}=2.0\,\mathrm{eV}$, comparing Exact
    (black solid) with SCBA (red dashed).  Columns are $S=0.5$, $1$, and $2$;
    rows are an isolated molecule and cavity ensembles with
    $N=1,2,4,$ and $8$.  Here $\nu_0=20\,\mathrm{meV}$, $T=0$,
    $\eta=0.5\,\mathrm{meV}$, and $g_\mathrm{c}=35.36\,\mathrm{meV}$ is
    fixed.  The isolated reference is the analytic Franck--Condon progression;
    cavity references use permutation-symmetric exact diagonalization.  Each
    SCBA curve is matched to the exact integrated area over
    $1.8$--$2.3\,\mathrm{eV}$; both curves use the exact peak as their common
    vertical scale.
    }
    \label{fig:discrete_cavity_zero_zero}
\end{figure}

\begin{figure}[!htbp]
    \centering
    \includegraphics[width=0.95\linewidth]{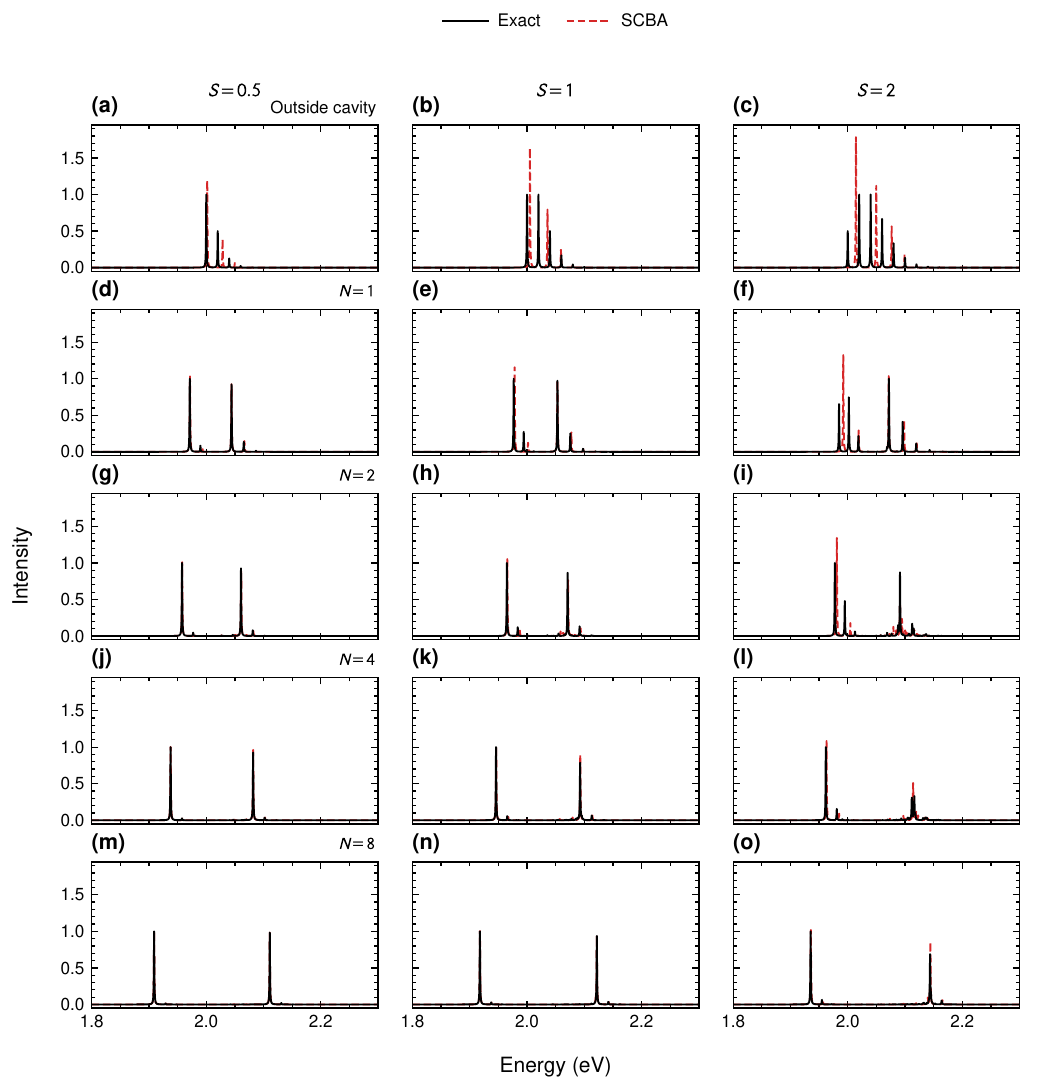}
    \caption{%
    Discrete-mode benchmark at vertical-transition resonance,
    $\omega_\mathrm{c}=\omega_0=\omega_{00}+S\nu_0=2.01$, $2.02$, and
    $2.04\,\mathrm{eV}$ across the three columns.  Methods, columns, rows, and
    common parameters are as in Fig.~\ref{fig:discrete_cavity_zero_zero}.  Each SCBA curve is matched
    to the exact integrated area over $1.8$--$2.3\,\mathrm{eV}$; both curves
    use the exact peak as their common vertical scale.
    }
    \label{fig:discrete_cavity_vertical}
\end{figure}

Figures~\ref{fig:discrete_cavity_zero_zero} and
\ref{fig:discrete_cavity_vertical} compare the same ensembles at two
cavity tunings.  Each molecule has one local mode with
$\nu_0=20\,\mathrm{meV}$ at $T=0$.

\FloatBarrier

The columns correspond to $\lambda=S\nu_0=10$, $20$, and
$40\,\mathrm{meV}$ and $g=\nu_0\sqrt S=14.14$, $20.00$, and
$28.28\,\mathrm{meV}$.  The common isolated-molecule row reproduces
Eq.~\ref{eq:SI_single_boson_A}; the cavity rows show how $S$, $N$, and the
tuning condition redistribute polariton and vibronic weight as
$\sqrt N g_\mathrm{c}$ grows.

At $T=0$, Eqs.~\ref{eq:SI_Gloc} and \ref{eq:SI_SCBA} reduce to
$\Sigma^R(\omega)=S\nu_0^2\bar G^R(\omega-\nu_0)$, with
$\bar G^R=[G_{BB}^R+(N-1)G_D^R]/N$ inside the cavity; the plotted molecular response is $-\operatorname{Im}G_{BB}^R$.
Outside the cavity this becomes the scalar isolated-molecule equation.
SCBA is closest to the leading exact features at the smallest $S$ and shifts
or misweights vibronic replicas as $S$ grows.  Changing $N$ changes both the
collective splitting and the number of dark-state channels; deviations are
nonmonotonic at small $N$ and generally smaller for the dominant polariton
peaks at the largest displayed $N$.

\subsection{Frenkel-Dimer Benchmarks}
\label{sec:SI_dimer_validity}

\paragraph{Model and site-local self-energy.}
\label{sec:dimer_model}
We next consider a symmetric one-excitation Frenkel dimer without a cavity,
\begin{align}
\hat H_{\rm dim}={}&
\omega_0\sum_{i=1}^{2}\hat d^\dagger_i \hat d_i
+J\bigl(\hat d^\dagger_1 \hat d_2 + \hat d^\dagger_2 \hat d_1\bigr)
+\sum_{i\alpha}\omega_\alpha
\hat b_{\alpha i}^\dagger\hat b_{\alpha i}
+\sum_i \hat d^\dagger_i \hat d_i\sum_\alpha c_\alpha
\bigl(\hat b_{\alpha i}+\hat b_{\alpha i}^\dagger\bigr).
\label{eq:SI_dimer_H}
\end{align}
For equal parallel transition dipoles, $\hat \mu \propto \hat B + \hat B^\dagger$, where $\hat B = (\hat d_1 + \hat d_2) / \sqrt{2}$.  The dark-state operator is $\hat D = (\hat d_1 - \hat d_2) / \sqrt{2}$.  The bright and dark states are eigenstates of the electronic Hamiltonian, with energies $E_{B,D}=\omega_0\pm J$.
Thus $J>0$ places the bright state above the dark state, whereas $J<0$ places it below, and the absorption is proportional to $-\operatorname{Im}G_B^R$.
Under SCBA, identical independent baths give a common site-local self-energy,
\begin{equation}
G_{B,D}^R(\omega)=
\left[\omega-\omega_0\mp J-\Sigma^R(\omega)+i\eta\right]^{-1},
\qquad
\bar G^R(\omega)=\frac{G_B^R(\omega)+G_D^R(\omega)}{2}.
\label{eq:SI_dimer_GBD}
\end{equation}
The continuous-bath kernel is already given in Eq.~\ref{eq:SI_SCBA_integral}; for the dimer, its internal local line is the $\bar G^R$ in Eq.~\ref{eq:SI_dimer_GBD}.
For a single mode of frequency $\nu_0$ at $T=0$, this reduces to $\Sigma^R(\omega)=g^2\bar G^R(\omega-\nu_0)$.

\paragraph{Single-mode scan and representative spectra.}
\label{sec:dimer_single_mode_scan}
With one undamped mode per site, we fix $|J|=25\,\mathrm{meV}$, $T=0$,
$\eta/\nu_0=0.005$, and $S=0.02$, $0.05$, or $0.10$, and vary the
vibrational frequency according to
\begin{equation}
r=\frac{2|J|}{\nu_0},\qquad
\nu_0=\frac{2|J|}{r},\qquad
g=\nu_0\sqrt S,\qquad
\lambda=S\nu_0.
\label{eq:SI_dimer_frequency_scan}
\end{equation}
Thus $|J|$ and $S$ are fixed while $\nu_0$, $g$, $\lambda$, and $\eta$ vary.
We compare exact diagonalization (ED) with SCBA.  For
$X\in\{\mathrm{ED},\mathrm{SCBA}\}$, let
$E_\pm=E_B\pm18\nu_0$ and define
\begin{equation}
\widetilde A_X(E)=\frac{A_X(E)}{\int_{E_-}^{E_+}dE'\,A_X(E')},\qquad
F_X(E)=\int_{E_-}^{E}dE'\,\widetilde A_X(E'),\qquad
W_1=\int_{E_-}^{E_+}dE\,\left|F_{\rm ED}(E)-F_{\rm SCBA}(E)\right|,
\label{eq:SI_dimer_W1}
\end{equation}
The density plotted against $x=(E-E_B)/\nu_0$ is
$p_X(x)=\nu_0\widetilde A_X(E_B+\nu_0x)$ and has unit area on
$x\in[-18,18]$.  We use the dimensionless error
\begin{equation}
w=\frac{W_1}{\nu_0}.
\label{eq:SI_dimer_dimensionless_W1}
\end{equation}
This dimensionless error measures differences in peak positions and spectral
weights.  The $J=0$ reference values are $0.00100$, $0.00591$, and $0.02150$
for $S=0.02$, $0.05$, and $0.10$.

\begin{figure}[htbp]
    \centering
    \includegraphics[width=0.96\linewidth]{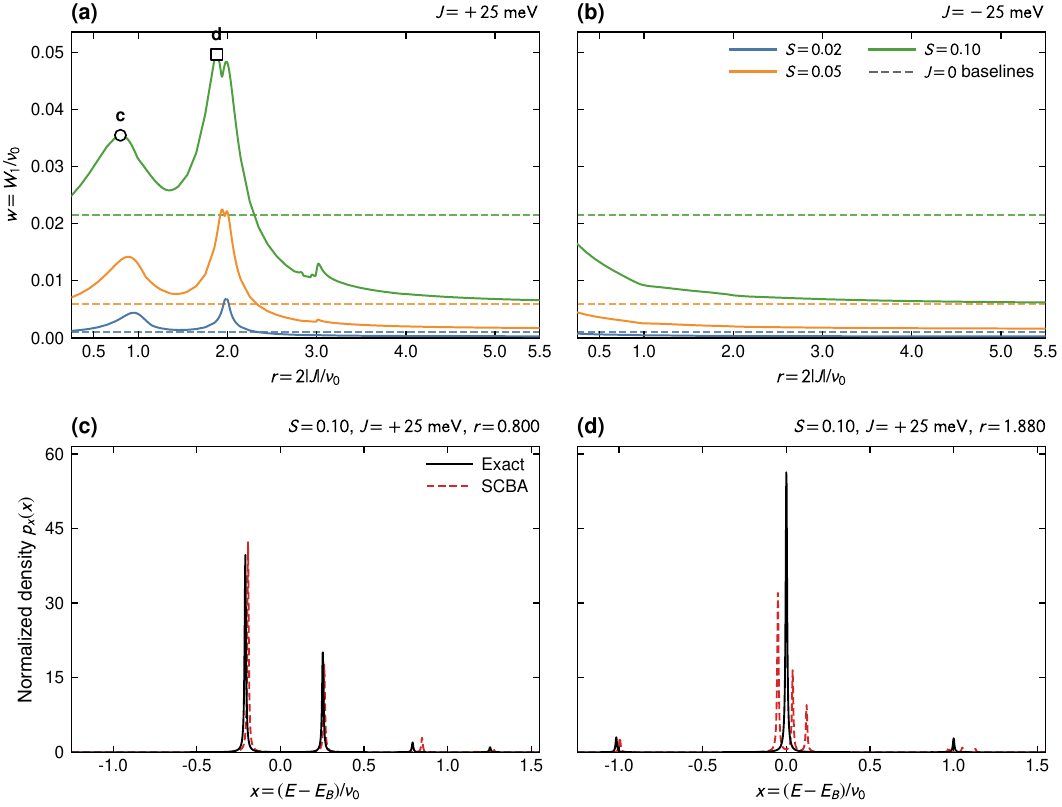}
    \caption{%
    Discrete-mode Frenkel-dimer benchmark at $|J|=25\,\mathrm{meV}$, $T=0$,
    and $\eta/\nu_0=0.005$.  (a,b) Dimensionless ED--SCBA distance
    $w=W_1/\nu_0$ versus $r=2|J|/\nu_0$ for $J=+25$ and $-25\,\mathrm{meV}$,
    respectively.  Colors denote $S=0.02$, $0.05$, and $0.10$; matching thin
    dashed lines are the $J=0$ references.  Open markers on the
    $S=0.10$ curve in (a) identify the spectra in (c,d).  (c,d) Exact (black
    solid) and SCBA (red dashed) spectra for $S=0.10$ and
    $J=+25\,\mathrm{meV}$: (c) $r=0.800$, near the first positive-$J$
    feature; (d) the central spectral window at $r=1.880$, where SCBA
    redistributes the dominant Exact line among nearby peaks.  Every spectrum
    is normalized to unit area on $x=(E-E_B)/\nu_0\in[-18,18]$; the narrower
    windows in (c,d) affect display only.  Scan curves connect calculated
    values without smoothing or interpolation.
    }
    \label{fig:dimer_single_mode_error}
\end{figure}

For $J>0$, the optically prepared bright state lies above the dark state, so
phonon emission can reach lower vibronic dark partners at $T=0$.  For $J<0$
the bright state lies below the dark state, making that on-shell emission
process energetically inaccessible, although virtual vibrational dressing
remains.  This ordering produces the contrasting scans in
Fig.~\ref{fig:dimer_single_mode_error}(a,b).

With $\hat b_\pm=(\hat b_1\pm\hat b_2)/\sqrt2$, the common mode is displaced
and the nontrivial difference-mode sector is
\begin{equation}
\hat H_-=J\tau_z+\nu_0\hat b_-^\dagger\hat b_-
+\frac{g}{\sqrt2}\tau_x(\hat b_-+\hat b_-^\dagger),
\qquad
\tau_z=|B\rangle\langle B|-|D\rangle\langle D|,
\quad
\tau_x=|B\rangle\langle D|+|D\rangle\langle B|.
\label{eq:SI_dimer_difference_mode}
\end{equation}
It conserves
\begin{equation}
\Pi=\tau_z(-1)^{\hat b_-^\dagger\hat b_-},
\qquad [\hat H_-,\Pi]=0.
\label{eq:SI_dimer_parity}
\end{equation}
For an occupation eigenvalue $n_-$, $|B,0\rangle$ has parity $+1$ and
$|D,n_-\rangle$ has parity $-(-1)^{n_-}$.  The bright source therefore shares
parity with dark partners of odd difference-mode occupation.  At weak
coupling their degeneracies satisfy
\begin{equation}
2J\simeq(2\ell+1)\nu_0,\qquad J>0,\qquad \ell=0,1,2,\ldots .
\label{eq:SI_dimer_odd_guides}
\end{equation}
This parity statement concerns the difference mode of the symmetric dimer, not an
arbitrary total phonon count across the two local modes or the full polariton
dark manifold studied in Sec.~\ref{sec:SI_bright_dark_degeneracy}.

The spectra identify the deviations behind the positive-$J$ scan.  Near
$r=0.800$, SCBA shifts the bright-state branches and changes their weights.
Near $r\simeq2$, it distributes the dominant central ED line among several
nearby peaks, and that redistribution supplies most of the integrated error
at this point.  It is an error in the calculated spectral structure rather
than an even-$n_-$ bright--dark avoided crossing.

\FloatBarrier
\paragraph{Continuous-bath comparison with HEOM.}
\label{sec:dimer_drude}
Each site in the continuous-bath comparison has an independent Drude--Lorentz
bath with $\lambda=30\,\mathrm{meV}$, correlation-rate scale
$\omega_f=24.8\,\mathrm{meV}$, and $T=300\,\mathrm{K}$.  SCBA follows the
HEOM spectra across all nine signed-$J$ values; the remaining line-shape
differences are small and nonmonotonic.  Thermal absorption and emission also
reduce the strong sign asymmetry of the zero-temperature single-mode case.

\begin{figure}[!ht]
    \centering
    \includegraphics[width=0.80\linewidth]{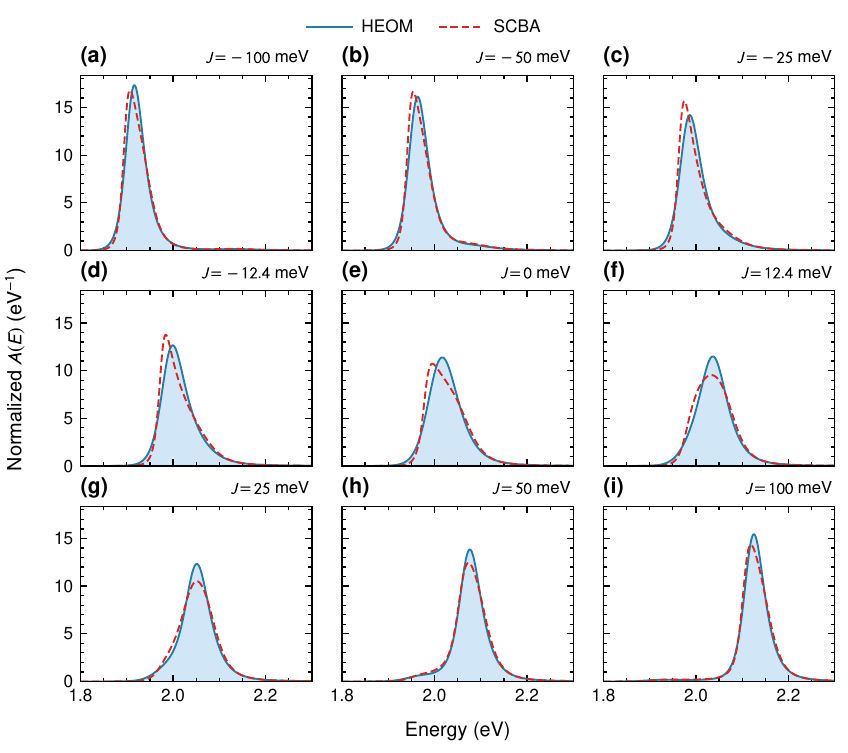}
    \caption{%
    Frenkel-dimer Drude--Lorentz benchmark against HEOM for $\lambda=30\,\mathrm{meV}$ per site, $\omega_f=24.8\,\mathrm{meV}$, and $T=300\,\mathrm{K}$.
    Panels (a)--(i) show the equal-parallel-dipole absorption for $J=-100,-50,-25,-12.4,0,+12.4,+25,+50,$ and $+100\,\mathrm{meV}$, respectively: HEOM is blue solid with light-blue shading and SCBA is red dashed.
    Each spectrum has unit area over $0.8000$--$3.1998\,\mathrm{eV}$; small
    negative HEOM Fourier artifacts are clipped.
    }
    \label{fig:dimer_drude_benchmark}
\end{figure}

\FloatBarrier
\paragraph{Included and omitted diagrams.}
\label{sec:dimer_diagrammatics}
The leading $B\to D\to B$ Fan process is included in BA and SCBA, and SCBA
iteration generates nested noncrossing rainbows.  Crossed diagrams and vertex
corrections at $O(g^4)$ are omitted.  The total ED--SCBA difference does not
isolate any one omitted contribution; that requires a diagram-resolved
calculation.
Agreement is best at small per-site $S$ and away from symmetry-allowed
vibronic degeneracies.

\begin{figure}[!ht]
    \centering
    \includegraphics[width=0.75\linewidth]{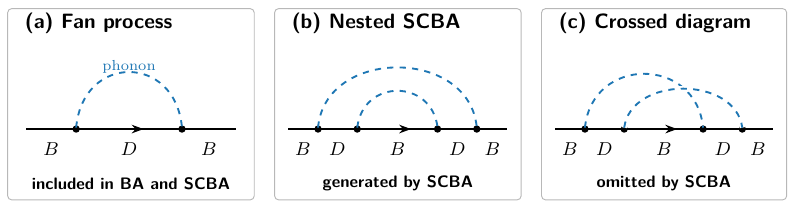}
    \caption{%
    Dimer diagrams for the site-local SCBA.
    Solid directed lines are matter propagators, blue phonon contractions connect the matter--phonon vertices, and the labels give the intermediate $B$ and $D$ states.
    (a) The leading $B\to D\to B$ Fan process is included in both BA and SCBA.
    (b) SCBA self-consistency generates the nested noncrossing $B\to D\to B\to D\to B$ rainbow.
    (c) The lowest crossed contraction has the same sequence of matter sectors but is omitted by SCBA.
    Unlike Fig.~\ref{fig:SI_Fan_diagrams}, this diagram labels the dimer's
    bright and dark intermediate states; it does not determine the numerical
    size of an omitted contribution.
    }
    \label{fig:dimer_diagrammatics}
\end{figure}

%%%%%%%%%%%%%%%%%%%%%%%%%%%%%%%%%%%%%%%%%%

\clearpage
\section{Numerical Implementation and Validation}
\label{sec:numerical_details}

\subsection{Energy Conventions and Calculation Settings}
\label{sec:energy_convention}

The molecular transition convention is the same in every calculation.  The
reorganization energy is defined by
\begin{equation}
\lambda=\frac{1}{\pi}\int_0^\infty d\nu\,\frac{J(\nu)}{\nu},
\label{eq:SI_reorganization}
\end{equation}
and the electronic diagonal in $\hat H_{\rm S}$ is the vertical transition,
\begin{equation}
\omega_0\equiv\omega_{00}+\lambda.
\label{eq:SI_vertical}
\end{equation}
For a discrete mode, $\lambda=S\nu_0$ and $c=\sqrt S\,\nu_0$, so the
displaced-oscillator poles are
\begin{equation}
E_n=\omega_0-\frac{c^2}{\nu_0}+n\nu_0
=\omega_{00}+n\nu_0.
\label{eq:SI_FC_poles}
\end{equation}
The SCBA, HEOM, MF-HEOM, and exact calculations retain the corresponding
unshifted linear bath coupling, and no spectrum is shifted after it is
calculated.  Main-text Drude calculations use vertical tuning,
$\omega_{\rm c}=\omega_0=2.03\,\mathrm{eV}$, unless a detuning is stated.  In
the discrete-mode benchmarks $\omega_{00}=2.0\,\mathrm{eV}$ and
$\nu_0=20\,\mathrm{meV}$; Fig.~\ref{fig:discrete_cavity_zero_zero} uses 0--0 tuning
($\omega_{\rm c}=\omega_{00}$), whereas Fig.~\ref{fig:discrete_cavity_vertical} uses vertical
tuning ($\omega_{\rm c}=\omega_0$).  Both hold the single-molecule coupling
at $g_{\rm c}=35.36\,\mathrm{meV}$, rather than holding
$\sqrt N g_{\rm c}$ fixed.

Table~\ref{tab:SI_numerical_settings} lists the grids used to solve the
self-consistency equations.  Continuous-bath results are interpolated only
after convergence to each figure's output grid.  The exact discrete-cavity
spectra use $1.8$--$2.3\,\mathrm{eV}$ at $0.05\,\mathrm{meV}$ spacing, and
the saved Drude-dimer spectra use $0.8$--$3.2\,\mathrm{eV}$ at
$0.2\,\mathrm{meV}$ spacing.

\begin{table}[htbp]
\caption{Numerical settings.  Grid ranges are in eV and spacings and
half-widths are in meV.  For the discrete dimer, $x$,
$\Delta x$, and $\eta/\nu_0$ are dimensionless.  The tangent-spaced bath sum
uses $N_{\rm b}=1000$ modes.}
\label{tab:SI_numerical_settings}
\small
\setlength{\tabcolsep}{3pt}
\begin{tabular}{@{}lllll@{}}
\hline
\parbox[t]{0.28\linewidth}{\raggedright Calculation} &
\parbox[t]{0.18\linewidth}{\raggedright Numerical method} &
\parbox[t]{0.13\linewidth}{\raggedright Grid range} &
\parbox[t]{0.12\linewidth}{\raggedright Grid spacing} &
\parbox[t]{0.13\linewidth}{\raggedright Half-width $\eta$}\\
\hline
\parbox[t]{0.28\linewidth}{\raggedright Cavity absorption, fixed $g_{\rm c}$; $N=1$ diagnostic}
 & \parbox[t]{0.18\linewidth}{\raggedright FFT convolution} & $[-1,4]$ & $0.050$ & $0.10$\\
\parbox[t]{0.28\linewidth}{\raggedright Collective linewidths, fixed $\sqrt N g_{\rm c}$; both baths, $N\leq64$}
 & \parbox[t]{0.18\linewidth}{\raggedright Linear FFT convolution} & $[-1,4]$ & $0.010$ & $0.01$\\
\parbox[t]{0.28\linewidth}{\raggedright Collective linewidths, fixed $\sqrt N g_{\rm c}$; both baths, $N=128$}
 & \parbox[t]{0.18\linewidth}{\raggedright Linear FFT convolution} & $[-1,4]$ & $0.0005$ & $0.001$\\
\parbox[t]{0.28\linewidth}{\raggedright Detuning scans}
 & \parbox[t]{0.18\linewidth}{\raggedright FFT convolution} & $[-1,4]$ & $0.010$ & $0.01$\\
\parbox[t]{0.28\linewidth}{\raggedright Single molecule, Brownian bath}
 & \parbox[t]{0.18\linewidth}{\raggedright FFT convolution} & $[-20,4.4]$ & $0.025$ & $0.10$\\
\parbox[t]{0.28\linewidth}{\raggedright Dimer, Drude bath}
 & \parbox[t]{0.18\linewidth}{\raggedright FFT convolution} & $[-1,4]$ & $0.050$ & $0.10$\\
\parbox[t]{0.28\linewidth}{\raggedright UP--dark gap}
 & \parbox[t]{0.18\linewidth}{\raggedright Discrete bath sum} & $[0.8,3.2]$ & $0.600$ & $0.01$\\
\parbox[t]{0.28\linewidth}{\raggedright Cavity, discrete vibrations}
 & \parbox[t]{0.18\linewidth}{\raggedright Frequency recursion} & $[1.5,3.2]$ & $0.050$ & $0.50$\\
\parbox[t]{0.28\linewidth}{\raggedright Dimer, discrete vibrations}
 & \parbox[t]{0.18\linewidth}{\raggedright Frequency recursion} & $x\in[-32,24]$ & $\Delta x=0.002$ & $0.005\nu_0$\\
\hline
\end{tabular}
\end{table}
For the collective-linewidth rows, the final data therefore use an explicit
$N$-dependent numerical broadening, not a uniform-$\eta$ data set.  The
$N=128$ spectra are evaluated on a $0.005\,\mathrm{meV}$ output grid after
self-consistency; finer output grids are used below only to test the direct
half-height extraction.
\FloatBarrier

\subsection{Continuous-Bath Convolution and SCBA Iteration}
\label{sec:SI_SCBA_numerics}

For the continuous Drude calculations, the emission and absorption kernels
in the Fan integral are
\begin{equation}
K_{\rm em}(\nu)=J(\nu)[\overline n(\nu)+1],\qquad
K_{\rm abs}(\nu)=J(\nu)\overline n(\nu).
\label{eq:SI_convolution_kernels}
\end{equation}
On the uniform electronic grid $\omega_j$, these kernels form two one-sided
Volterra convolutions with the local retarded propagator $\bar G^R$.  They are
evaluated by FFT after zero padding to at least $2N_\omega-1$ points, thereby
forming a linear rather than periodic convolution.  Trapezoidal half weights
are applied at $\nu=0$ and the finite electronic-grid boundaries.  In the
plotted frequency region, the omitted high-frequency Drude emission tail is
restored from the analytic large-$\nu$ expansion of $G_0^R$; the thermal
absorption tail is exponentially suppressed.

Evaluating Eq.~\ref{eq:SI_SCBA_integral} with the vibration-free local
propagator gives the single-shot BA self-energy, which initializes the SCBA
iteration.  At iteration $m$, the same self-energy update evaluated with
$\bar G^{R,(m)}$ gives the proposed value
$\widetilde\Sigma^{R,(m+1)}$.  Adaptive mixing forms
\begin{equation}
\Sigma^{R,(m+1)}(\omega_j)
=(1-\xi_m)\Sigma^{R,(m)}(\omega_j)
+\xi_m\widetilde\Sigma^{R,(m+1)}(\omega_j).
\label{eq:SI_mixing}
\end{equation}
Before mixing, convergence is measured by the maximum difference between the
proposed and preceding self-energies over the full calculation grid,
\begin{equation}
\epsilon_\Sigma^{(m)}=\max_j\left|
\widetilde\Sigma^{R,(m+1)}(\omega_j)
-\Sigma^{R,(m)}(\omega_j)\right|
<10^{-8}\,E_{\rm h}\;(=2.72\times10^{-7}\,\mathrm{eV}).
\label{eq:SI_global_residual}
\end{equation}
Mixing begins at $\xi_m=1$; if the residual exceeds the preceding residual by
more than 25\%, $\xi_m$ is halved, down to $0.2$.  At most 200 updates are
allowed.  Exhausting that limit raises a failure and is not counted as
convergence.\\

\noindent\textbf{SCBA iteration.}
\begin{enumerate}
\setlength{\itemsep}{1pt}
\setlength{\parskip}{0pt}
\setlength{\parsep}{0pt}
\item Construct the electronic grid and the two kernels in
 Eq.~\ref{eq:SI_convolution_kernels}; evaluate the single-shot BA self-energy
 $\Sigma^{R,(0)}$.
\item Construct the local retarded propagator $\bar G^{R,(m)}$ from the
current $\Sigma^{R,(m)}$.
\item Apply Eq.~\ref{eq:SI_SCBA_integral} by linear convolution to obtain
$\widetilde\Sigma^{R,(m+1)}$.
\item Evaluate the unmixed full-grid residual in
Eq.~\ref{eq:SI_global_residual}, adjust $\xi_m$ when required, and form the
update in Eq.~\ref{eq:SI_mixing}.
\item Repeat steps 2--4 until the residual is below tolerance, with a hard
failure after 200 updates.  Interpolate the converged complex self-energy to
the requested output grid and construct the absorption spectrum.
\end{enumerate}

The near-degeneracy scan uses instead the independent tangent discretization
of the bath~\cite{Wang_Thoss_2006,Wang_Thoss_2007},
\begin{align}
\omega_\alpha&=\omega_f\tan\!\left[\frac{\pi}{2}
\left(1-\frac{\alpha}{N_{\rm b}+1}\right)\right],
\quad \alpha=1,\ldots,N_{\rm b},\nonumber\\
c_\alpha&=\sqrt{\frac{\lambda\omega_\alpha}{N_{\rm b}+1}}.
\label{eq:SI_tangent}
\end{align}
This quadrature supplies the independent comparison below; the other
continuous-bath curves use FFT convolution.

\subsection{Convergence and Numerical Broadening}
\label{sec:SI_convergence}

\begin{figure}[htbp]
\centering
\includegraphics[width=1.0\linewidth]{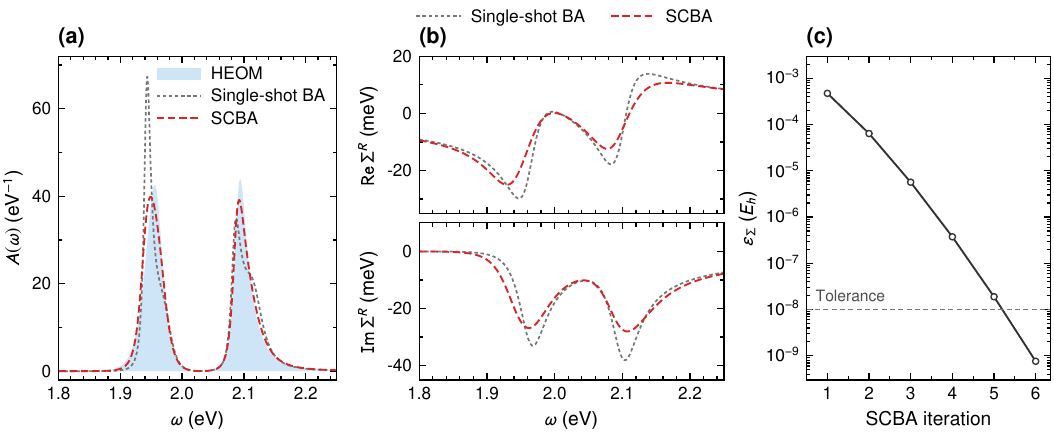}
\caption{BA and SCBA at vertical resonance
($\omega_{\rm c}=\omega_0=2.03\,\mathrm{eV}$, $N=1$,
$g_{\rm c}=68.1\,\mathrm{meV}$, and $\kappa=0$).
(a) Absorption: HEOM (light-blue shading), single-shot BA (gray short-dashed),
and SCBA (red dashed).  Each comparison curve is matched to the HEOM
area over their shared energy interval.
(b) Frequency-dependent real (upper axis) and imaginary (lower axis) parts
of $\Sigma^R$, with aligned axes and a shared frequency coordinate; the
method encodings are the same as in panel~(a).
(c) The saved full-grid unmixed residual $\epsilon_\Sigma$ versus SCBA update, together with
the $10^{-8}\,E_{\rm h}$ stopping threshold.  The final value is
$7.6006\times10^{-10}\,E_{\rm h}$.
HEOM data are reproduced from Ref.~\citenum{Ying_mn24}.}
\label{fig:SI_SCBA_diagnostics}
\end{figure}

\paragraph{One molecule in a cavity.}
\label{sec:SI_N1_convergence}
Figure~\ref{fig:SI_SCBA_diagnostics} shows the $N=1$ calculation at vertical
resonance, with $g_{\rm c}=68.1\,\mathrm{meV}$ and $\kappa=0$.  This example
also connects to earlier single-emitter cavity-QED studies based on exact
diagonalization and polaron-frame SCBA~\cite{Kaer2013,Hornecker2017}.  With
$\xi_m=1$, six SCBA updates reduce the full-grid unmixed residual below the
$10^{-8}\,E_{\rm h}$ tolerance; Fig.~\ref{fig:SI_SCBA_diagnostics}(c) gives
the complete history.

\paragraph{Iteration, causality, and grid checks.}
All 233 archived SCBA parameter sets underlying this diagnostic, the
main-text collective-linewidth and LP-detuning results,
Fig.~\ref{fig:SI_detuning_additional}, and
Fig.~\ref{fig:collective_limit_comparison} converged in
5--20 updates and satisfied the retarded-self-energy causality condition
$\operatorname{Im}\Sigma^R(\omega)\leq0$.  The twelve cases in
Fig.~\ref{fig:single_site_drude_benchmark} also satisfy this condition and converge in 55--72
fixed-mixing updates.  Table~\ref{tab:convolution} reports grid-refinement
tests for four representative spectra: no peak moves on the output grid, the
largest FWHM change is $7.2\times10^{-6}\,\mathrm{meV}$, and the unit-area
spectral $L^1$ difference is below $4.6\times10^{-7}$.

The 50 fixed-collective-coupling calculations retained for the present
half-width validation comprise 24 uniform-$\eta=0.01\,\mathrm{meV}$
production cases, ten same-$\eta$ grid refinements, ten
$\eta=0.005\,\mathrm{meV}$ cases spanning both fit windows and the two
$N=128$ endpoints, four $N=128$, $\eta=0.001\,\mathrm{meV}$ cases on two
internal grids, and two $N=128$, $\eta=0.0005\,\mathrm{meV}$ cases.  They
converge in 5--20 updates, have final full-grid residuals no larger than
$2.70\times10^{-4}\,\mathrm{meV}$, below the
$2.72\times10^{-4}\,\mathrm{meV}$ stopping threshold, satisfy
$\operatorname{Im}\Sigma^R\leq0$, and yield positive spectra with reliable
LP and UP half-height crossings.

\begin{table}[htbp]
\caption{Grid-refinement tests for four representative FFT-convolution
calculations.  Entries are maxima over resolved branches;
$\Delta_1=\int d\omega\,|\widehat A_{\rm fine}-\widehat A_{\rm coarse}|$
for unit-area spectra, and all energy columns are in meV.}
\label{tab:convolution}
\begin{ruledtabular}
\begin{tabular}{lccc}
Calculation & $\max|\Delta\Gamma|$ & $\Delta_1$ & $\max|\Delta\Sigma|$\\
\hline
$N=1$ cavity diagnostic & $7.10\times10^{-6}$ & $8.06\times10^{-8}$ & $6.04\times10^{-6}$\\
Slow bath, $N=128$ & $4.69\times10^{-6}$ & $4.54\times10^{-7}$ & $4.12\times10^{-5}$\\
Detuning sweep, $N=64$ & $6.75\times10^{-8}$ & $2.82\times10^{-9}$ & $1.51\times10^{-5}$\\
Uncoupled fast-bath molecule & $5.71\times10^{-6}$ & $8.04\times10^{-8}$ & $5.99\times10^{-6}$\\
\end{tabular}
\end{ruledtabular}
\end{table}

With $N_{\rm b}=2000$ and 8000 electronic frequencies over
$-1$--$4\,\mathrm{eV}$, the tangent-grid comparison changes peak
positions by at most $0.025\,\mathrm{meV}$, FWHM values by at most
$0.039\,\mathrm{meV}$, and unit-area spectra by at most
$3.94\times10^{-3}$ in $L^1$ distance.  For the uncoupled molecule, the
near-degeneracy tangent calculation gives $88.146\,\mathrm{meV}$, whereas the
FFT calculation gives $88.349\,\mathrm{meV}$.  The $0.203\,\mathrm{meV}$
difference combines method, grid, and broadening choices and is not assigned
to $\eta$ alone.

For the near-degeneracy calculation at $\delta_{+D}=20\,\mathrm{meV}$ and
$N=4$ or $64$, doubling the bath or frequency grid, widening the calculated
range, halving $\eta$, tightening the stopping tolerance, and varying the
fixed mixing fraction change the UP FWHM by at most
$0.0244\,\mathrm{meV}$.  At gaps of $0.02$--$0.5\,\mathrm{meV}$, four
initializations change it by at most $1.41\times10^{-4}\,\mathrm{meV}$.

The energy convention was checked independently by diagonalizing an
80-state displaced oscillator for $S=0,0.5,1,$ and $2$: the first six poles
agree with Eq.~\ref{eq:SI_FC_poles} to $4.5\times10^{-16}\,\mathrm{eV}$.
For the representative Drude case $\lambda=30\,\mathrm{meV}$ and
$\omega_f=24.8\,\mathrm{meV}$, the exact-cumulant and HEOM maxima both occur
at $2.016\,\mathrm{eV}$ on the common grid.

\paragraph{Finite retarded half-width.}
The half-width $\eta$ enters every molecular and photon propagator during
self-consistency, not as a final convolution; it is a numerical retarded
half-width and is distinct from the physical cavity loss and the
bath-induced linewidth.  The final fixed-collective-coupling scan uses
$\eta=0.01\,\mathrm{meV}$ and $\Delta\omega=0.01\,\mathrm{meV}$ for
$N\leq64$, but uses $\eta=0.001\,\mathrm{meV}$ and
$\Delta\omega=0.0005\,\mathrm{meV}$ for $N=128$ in both baths.  Repeating
the slow-bath fit points and $N=128$, and all five fast-bath fit points, at
$\eta=0.01\,\mathrm{meV}$ with $\Delta\omega=0.005\,\mathrm{meV}$ changes
any LP FWHM by at most $2.15\times10^{-5}\,\mathrm{meV}$ and changes the
slow- and fast-bath fitted exponents by $2.90\times10^{-7}$ and
$8.80\times10^{-9}$, respectively.

This grid convergence does not establish the $\eta\to0$ limit.  At
$\eta=0.005\,\mathrm{meV}$ on the $0.005\,\mathrm{meV}$ grid, the slow- and
fast-bath fits give $k=-0.543223$ ($R^2=0.999891$) and
$k=-0.980005$ ($R^2=0.999802$).  Relative to $\eta=0.01\,\mathrm{meV}$,
the fast-bath widths change by $0.24\%$, $0.48\%$, $0.96\%$, $1.86\%$, and
$3.51\%$ at $N=8,16,32,64,128$, respectively.  The uniform
$\eta=0.005\,\mathrm{meV}$ fast exponent differs from the final mixed-$\eta$
value by only $0.00717$.

For $N=128$, halving the internal spacing from $0.001$ to
$0.0005\,\mathrm{meV}$ changes either LP FWHM by less than
$2\times10^{-9}$ relative.  At $\eta=0.001\,\mathrm{meV}$ the final slow-
and fast-bath widths are $2.712520$ and $0.267031\,\mathrm{meV}$.
Refining the output spacing from $0.005$ to $0.001\,\mathrm{meV}$ changes
them by $0.00024\%$ and $0.0369\%$, while the integrated bright spectral
weights over the saved windows remain stable to better than
$3\times10^{-11}$.  Reducing $\eta$ further to
$0.0005\,\mathrm{meV}$ gives $2.711850$ and
$0.266018\,\mathrm{meV}$, changes of $-0.0247\%$ and $-0.379\%$.
The corresponding fast exponent is $-0.988273$, only $0.00110$ below the
adopted value.  A linear endpoint-only $\eta\to0$ diagnostic gives
$k=-0.989353$; it is not used as production data, and no $2\eta$
subtraction or other post-processing correction is applied.

\subsection{Discrete-Mode Calculations and Convergence}

\paragraph{Cavity ensembles.}
For the outside-cavity rows of Figs.~\ref{fig:discrete_cavity_zero_zero} and
\ref{fig:discrete_cavity_vertical}, the exact
Franck--Condon reference is Eq.~\ref{eq:SI_single_boson_A}, summed over
$n=0,\ldots,60$; the two outside-cavity rows are consequently identical.
For the cavity rows, every molecule has an independent local oscillator and
ED is restricted exactly to the totally permutation-symmetric sector reached
by the normalized bright source.  The cutoff $Q$ in
Table~\ref{tab:SI_discrete_basis} is the total number of vibrational quanta
in that ED sector; it is distinct from the excitation number $Q_{\rm exc}$
and the pseudoparticle constraint $Q_{\rm PP}$.

\begin{table}[htbp]
\caption{Permutation-symmetric cavity-ensemble ED dimensions and solvers.
The calculations use $Q=28$; $Q=30$ tests convergence.}
\label{tab:SI_discrete_basis}
\begin{ruledtabular}
\begin{tabular}{rccc}
$N$ & Basis size ($Q=28$) & Basis size ($Q=30$) & Diagonalization method\\
\hline
1 & 58 & 62 & Dense\\
2 & 660 & 752 & Dense\\
4 & 9382 & 11940 & 5000-step Lanczos\\
8 & 58403 & 85073 & 5000-step Lanczos\\
\end{tabular}
\end{ruledtabular}
\end{table}

The $Q=30$ calculation changes the peak-normalized spectra by at most
$7.86\times10^{-8}$ across both tuning protocols.  For $N=4$ and $8$, the
Hermitian Lanczos tridiagonalization is evaluated through its projected
continued-fraction resolvent; reducing it to 4000 steps changes the spectra
by at most $2.49\times10^{-9}$.

At $T=0$, the corresponding discrete-mode SCBA is solved by an
ascending-frequency causal recursion on 34001 points from $1.5$ to
$3.2\,\mathrm{eV}$.  The $0.05\,\mathrm{meV}$ spacing makes the
$20\,\mathrm{meV}$ vibrational shift exactly 400 points, so this recursion
requires no iterative mixing.  The common half-width is
$\eta=0.5\,\mathrm{meV}$.  Repeating all cases at $0.10\,\mathrm{meV}$
spacing changes the peak-normalized displayed spectrum by at most
$9.88\times10^{-3}$.  The exact curves are evaluated on 10001 points over
$1.8$--$2.3\,\mathrm{eV}$ at $0.05\,\mathrm{meV}$ spacing.  The displayed
curves satisfy the fixed-point, retarded-causality, positivity, and
positive-area checks; the cavity results also satisfy the cutoff check.

\paragraph{Dimers.}
For the single-mode ED calculation, common and difference vibrational modes
are used.  The displaced common oscillator is treated analytically, while
the final Lehmann sum and difference-mode basis retain
$n_\pm=0,\ldots,N_{\rm vib}-1$.  Only the $2N_{\rm vib}$-dimensional
bright/dark difference-mode Hamiltonian is diagonalized.  The finite-$J$
curves use $N_{\rm vib}=10$, giving dimension 20 with 10 states in each
parity block.

SCBA and ED use $x=(E-E_B)/\nu_0\in[-32,24]$ with
$\Delta x=0.002$ (28001 points) and $\eta/\nu_0=0.005$.  The reported
$W_1$ is evaluated over $x\in[-18,18]$ from trapezoidal point weights,
equivalently as the integral of the absolute difference of the cumulative
distributions, without smoothing, interpolation, fitting, or peak
normalization.  At $T=0$, SCBA again uses a causal ascending-frequency
sweep, with absolute and relative fixed-point tolerances
$10^{-9}\,\mathrm{eV}$ and $10^{-8}$.

The finite-$J$ scan and the three displayed $J=0$ references satisfy the
fixed-point, retarded-causality, channel-sum, positivity, and raw-area checks;
the largest relative area mismatch is $5.83\times10^{-7}$.
Figure~\ref{fig:dimer_single_mode_error} uses these values together with 34
calculated midpoint/refined-$r$ values at the same physical and numerical
settings; points are connected in increasing $r$ without interpolation or
smoothing.
At eight positive- and negative-$J$ test points near $r\simeq1$ and $2$,
changing $\Delta x$ from $0.002$ to $0.001$ and $0.0005$ changes $w$ by at
most $2.80\times10^{-5}$ and $3.43\times10^{-5}$, respectively; a half-cell
displacement changes $w$ by at most $5.84\times10^{-5}$, whereas increasing
$N_{\rm vib}$ from 10 to 12 changes it by at most $2.28\times10^{-14}$.
The largest midpoint deviation from the chord joining its neighboring
sampled points is $4.61\times10^{-5}$.
Changing the metric support from $[-18,18]$ to $[-16,16]$ or $[-20,20]$
changes $w$ by at most $1.10\times10^{-6}$.
These variations are small relative to the broad ED--SCBA distance but
comparable to the smallest visible zigzags, which are therefore not assigned
physical significance.

\subsection{Spectral Normalization and Linewidth Extraction}
\label{sec:linewidth}

Table~\ref{tab:SI_normalization} collects the distinct plotting
normalizations.  A single area-matching factor always multiplies the full
comparison spectrum and therefore preserves its internal peak-height ratios.
Display offsets and common display factors are applied only after the stated
area operation.

\begin{table}[htbp]
\caption{Normalization conventions.  A ``shared energy interval'' is the
overlap of the saved frequency grids.  A narrower display window does not
change the normalization interval.}
\label{tab:SI_normalization}
\small
\setlength{\tabcolsep}{4pt}
\begin{tabular}{@{}lll@{}}
\hline
\parbox[t]{0.21\linewidth}{\raggedright Figure(s)} &
\parbox[t]{0.42\linewidth}{\raggedright Normalization} &
\parbox[t]{0.23\linewidth}{\raggedright Normalization interval}\\
\hline
\parbox[t]{0.21\linewidth}{\raggedright Main Fig.~2; Fig.~\ref{fig:SI_SCBA_diagnostics}}
 & \parbox[t]{0.42\linewidth}{\raggedright Match each BA/SCBA area to HEOM} & \parbox[t]{0.23\linewidth}{\raggedright Shared energy interval}\\
\parbox[t]{0.21\linewidth}{\raggedright Fig.~\ref{fig:SI_detuning_additional}}
 & \parbox[t]{0.42\linewidth}{\raggedright HEOM reference, then common scale/offset} & $1.5$--$2.5\,\mathrm{eV}$\\
\parbox[t]{0.21\linewidth}{\raggedright Fig.~\ref{fig:single_site_drude_benchmark}}
 & \parbox[t]{0.42\linewidth}{\raggedright Match each area to exact cumulant; exact peak scale} & \parbox[t]{0.23\linewidth}{\raggedright Shared calculated grid}\\
\parbox[t]{0.21\linewidth}{\raggedright Main Fig.~4; Figs.~S4 and S5}
 & \parbox[t]{0.42\linewidth}{\raggedright Match SCBA area to Exact; exact peak scale} & $1.8$--$2.3\,\mathrm{eV}$\\
\parbox[t]{0.21\linewidth}{\raggedright Fig.~\ref{fig:brownian_benchmark}}
 & \parbox[t]{0.42\linewidth}{\raggedright Each curve has unit area} & $1.6$--$3.8\,\mathrm{eV}$\\
\parbox[t]{0.21\linewidth}{\raggedright Fig.~\ref{fig:dimer_single_mode_error}}
 & \parbox[t]{0.42\linewidth}{\raggedright Each curve has unit area} & $x\in[-18,18]$\\
\parbox[t]{0.21\linewidth}{\raggedright Fig.~\ref{fig:dimer_drude_benchmark}}
 & \parbox[t]{0.42\linewidth}{\raggedright Each curve has unit area} & $0.8$--$3.1998\,\mathrm{eV}$\\
\parbox[t]{0.21\linewidth}{\raggedright Fig.~\ref{fig:collective_limit_comparison}}
 & \parbox[t]{0.42\linewidth}{\raggedright Separate unit areas, then one common display factor} & $1.6$--$2.4\,\mathrm{eV}$\\
\hline
\end{tabular}
\end{table}

For the main-text Fig.~2 pairs, integration over the common
$1.5$--$2.5\,\mathrm{eV}$ support gives raw HEOM areas
$0.998432\pi$--$0.999963\pi$ and raw SCBA areas
$0.998516\pi$--$0.999763\pi$.  The resulting SCBA-to-HEOM factors are
$0.9999159$--$1.0002005$; restricting the integral to the displayed interval
changes a factor by at most $5.33\times10^{-4}$.  These are finite-window
checks, not a claim that every plotted observable obeys the same unit-area or
absorbed-power sum rule.
For the collective-limit comparison in
Fig.~\ref{fig:collective_limit_comparison}, the raw areas over
$1.6$--$2.4\,\mathrm{eV}$ are $3.13604$--$3.13751$ (MF-HEOM) and
$3.13245$--$3.13392$ (SCBA); the SCBA-to-MF-HEOM factors are
$1.001145$--$1.001153$.

All reported SCBA, HEOM, and MF-HEOM linewidths use direct half-height
crossings rather than Lorentzian fits.  An 11-point cubic Savitzky--Golay
filter is used only to locate candidate peaks; the displayed spectra,
baselines, and crossings remain unsmoothed.  Candidate peaks use normalized
prominence 0.01 and 10-point separation, with 0.005 and five points as
fallbacks.  The two strongest peaks define LP and UP, and the minimum of the
original spectrum between them defines the separating valley.

For each branch, the baseline is the minimum between its outer frequency
bound and that valley.  Linear interpolation on the original grid gives the
two half-height crossings and their separation.  A width is rejected if a
crossing is absent, it spans no more than two grid intervals, or its
normalized prominence is below 0.005; values below 0.01 are also marked weak.

% The quoted numerical uncertainty is one frequency-grid interval, spanning $0.006$--$1.500\,\mathrm{meV}$ across these data sets; the upper value belongs to the additional low-cavity-frequency HEOM spectra. This is a crossing resolution estimate, not an error budget for finite $\eta$, method error, or model error.  Every reported linewidth passes the reliability criterion.

\subsection{Collective-Linewidth Fits}
\label{sec:fit_windows}

The fixed-collective-coupling scaling scan holds the collective coupling
$\sqrt N g_{\rm c}$ fixed.  Its fit intervals are assessed with
$\mathcal K_N=\omega_f/\sqrt{2(\lambda/N)k_{\rm B}T}$ at $T=300\,\mathrm K$
and $\lambda=30\,\mathrm{meV}$.  The selected slow-bath points have
$\mathcal K_N=0.063$--$0.126\ll1$.  The selected fast-bath points have
$\mathcal K_N=1.78$--$7.12$; all exceed unity, although the lowest is not
asymptotically large.  The windows are therefore described as approximate
slow- and fast-fluctuation regimes.  Fits to the selected SCBA points give
$k=-0.543196$ ($R^2=0.999891$) and
$k=-0.987177$ ($R^2=0.999938$), respectively.  For
$\ln(\Gamma_{\rm LP}/\mathrm{meV})=b+k\ln N$, the corresponding intercepts
are $b=4.152507$ and $3.472872$.

\begin{table}[htbp]
\caption{Kubo parameter, SCBA LP FWHM, and inclusion in the prescribed
fixed-collective-coupling fits.  Points with $N\leq64$ use
$\eta=\Delta\omega=0.01\,\mathrm{meV}$; both $N=128$ points use
$\eta=0.001\,\mathrm{meV}$ and $\Delta\omega=0.0005\,\mathrm{meV}$ on
$[-1,4]\,\mathrm{eV}$.  The slow and fast cutoff frequencies are $2.48$ and
$24.8\,\mathrm{meV}$.}
\label{tab:kubo}
\small
\begin{ruledtabular}
\begin{tabular}{rccc@{\quad}ccc}
&\multicolumn{3}{c}{Slow bath}&\multicolumn{3}{c}{Fast bath}\\
$N$&$\mathcal K_N$&FWHM (meV)&Used in fit&$\mathcal K_N$&FWHM (meV)&Used in fit\\
\hline
1 & 0.063 & 63.444 & yes & 0.630 & 32.692 & no \\
2 & 0.089 & 43.820 & yes & 0.891 & 17.006 & no \\
3 & 0.109 & 35.066 & yes & 1.091 & 11.316 & no \\
4 & 0.126 & 29.852 & yes & 1.259 & 8.440 & no \\
5 & 0.141 & 26.302 & no & 1.408 & 6.719 & no \\
6 & 0.154 & 23.688 & no & 1.542 & 5.578 & no \\
7 & 0.167 & 21.661 & no & 1.666 & 4.768 & no \\
8 & 0.178 & 20.031 & no & 1.781 & 4.163 & yes \\
16 & 0.252 & 13.162 & no & 2.519 & 2.072 & yes \\
32 & 0.356 & 8.354 & no & 3.562 & 1.045 & yes \\
64 & 0.504 & 4.994 & no & 5.038 & 0.537 & yes \\
128 & 0.712 & 2.713 & no & 7.124 & 0.267 & yes \\
\end{tabular}
\end{ruledtabular}
\end{table}

\FloatBarrier

\clearpage
\section{Additional Numerical Results}
\label{sec:SI_additional_detuning}

\subsection{Detuning-Dependent Spectra and Upper-Polariton Linewidths}
\label{sec:SI_detuning_spectra}

\begin{figure}[htbp]
\centering
\includegraphics[width=0.98\linewidth]{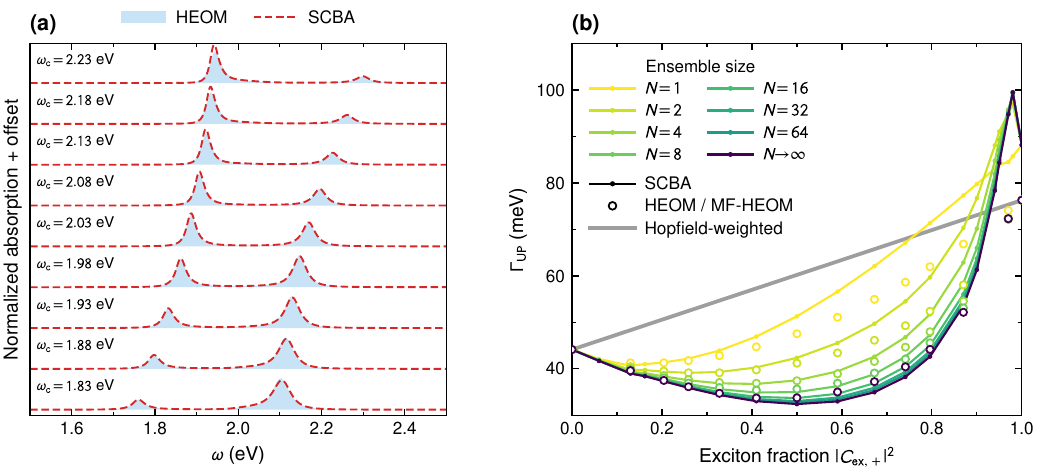}
\caption{Detuning-dependent results for independent Drude--Lorentz baths
($\lambda=30\,\mathrm{meV}$, $\omega_f=24.8\,\mathrm{meV}$,
$T=300\,\mathrm{K}$, $\omega_{00}=2.00\,\mathrm{eV}$,
$\omega_0=2.03\,\mathrm{eV}$) and cavity FWHM
$\kappa=44.15\,\mathrm{meV}$.  (a) Offset $N=4$ spectra for
$g_\mathrm{c}=68.1\,\mathrm{meV}$ and
$\omega_\mathrm{c}=1.83$--$2.23\,\mathrm{eV}$: HEOM (blue shading) and
SCBA (red dashed).  Each pair uses HEOM-reference area matching over
$1.5$--$2.5\,\mathrm{eV}$ followed by a common peak scale and offset.
(b) UP FWHM versus $|C_{\mathrm{ex},+}|^2$ for
$N=1,2,4,8,16,32,64$, and the collective limit, with
$g_\mathrm{c}=136.2\,\mathrm{meV}/\sqrt N$.  Solid curves are SCBA; open
points are HEOM ($N\leq8$) or MF-HEOM (collective limit); the gray line is
the Hopfield-weighted reference with
$\Gamma_\mathrm{ex}=76.28\,\mathrm{meV}$.  The separately calculated
pure-exciton endpoints are $76.28\,\mathrm{meV}$ (HEOM) and
$88.15\,\mathrm{meV}$ (SCBA).  Small-$N$ HEOM data are from
Ref.~\citenum{Ying_mn24}.}
\label{fig:SI_detuning_additional}
\end{figure}

At fixed collective coupling $\sqrt{N}g_\mathrm{c}=136.2\,\mathrm{meV}$,
the UP linewidth approaches the cavity loss on its photon-rich side, passes
through an $N$-dependent minimum, and then rises nonlinearly with exciton
fraction [Fig.~\ref{fig:SI_detuning_additional}(b)].  A Hopfield-weighted
linewidth does not reproduce this curvature or its ensemble-size dependence.
HEOM supplies the $N=1,2,4,$ and $8$ benchmarks and MF-HEOM the
collective-limit points.  SCBA overbroadens the strongly excitonic UP relative
to both; Sec.~\ref{sec:SI_bright_dark_degeneracy} resolves the corresponding
bright and dark internal channels.

The $N=4$ spectra in panel~(a) follow the anticrossing as
$\omega_\mathrm{c}$ increases from $1.83$ to $2.23\,\mathrm{eV}$.  SCBA
tracks the principal HEOM peaks and the exchange of their relative prominence.
For each pair, one constant matches the SCBA area to the raw HEOM area over
$1.5$--$2.5\,\mathrm{eV}$; the HEOM peak sets their shared display scale, and
both curves receive the same row offset.

\FloatBarrier
\subsection{Upper-Polariton Broadening near Dark-State Degeneracy}
\label{sec:SI_bright_dark_degeneracy}

\begin{figure}[!b]
\centering
\includegraphics[width=0.80\linewidth]{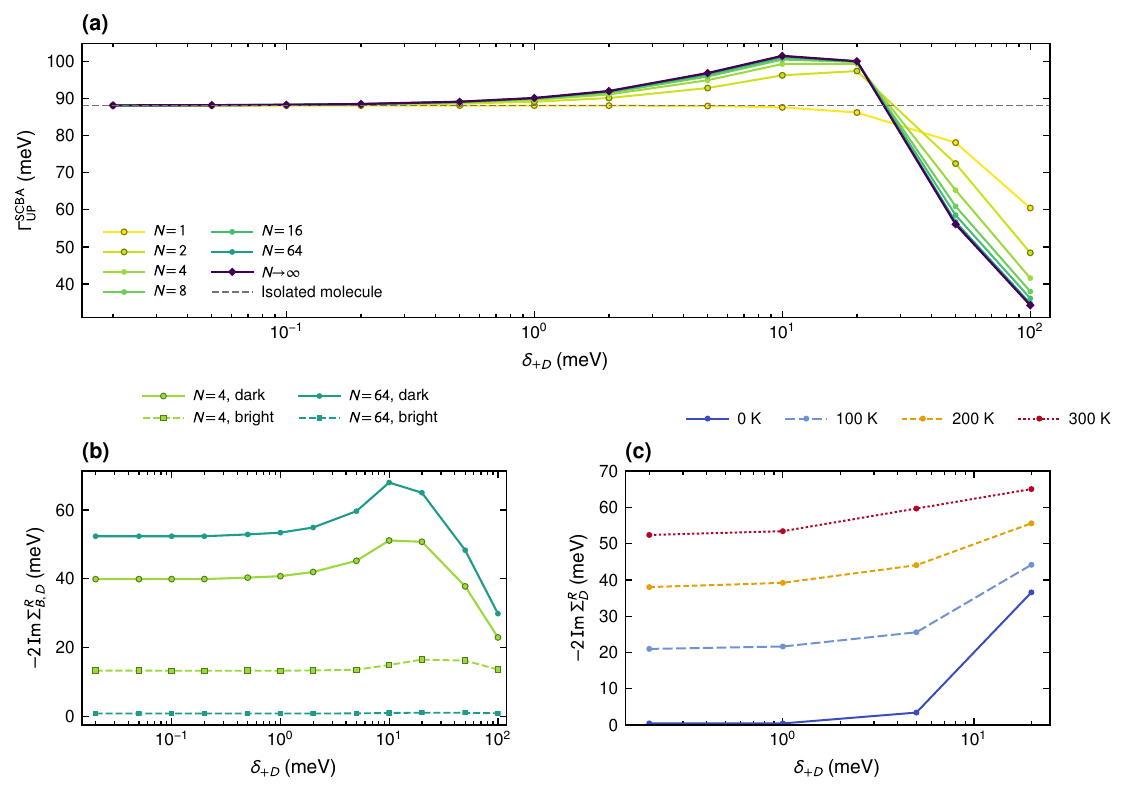}
\caption{SCBA behavior near UP--dark degeneracy for
$G=136.2\,\mathrm{meV}$, $\omega_0=2.03\,\mathrm{eV}$,
$\kappa=44.15\,\mathrm{meV}$, and a Drude--Lorentz bath with
$\lambda=30\,\mathrm{meV}$ and $\omega_f=24.8\,\mathrm{meV}$.
Panels~(a,b) use $T=300\,\mathrm K$.
(a) UP FWHM versus the unperturbed gap
$\delta_{+D}=\omega_+-\omega_0$ for $N=1,2,4,8,16,64$, and the collective
limit; colors follow Fig.~\ref{fig:SI_detuning_additional}, and the gray
dashed line is the tangent-grid isolated-molecule reference.  Here
$\omega_\mathrm{c}>0$ requires $\delta_{+D}>9.10\,\mathrm{meV}$; the
approach to the isolated-molecule linewidth below $1\,\mathrm{meV}$ is a
formal limiting behavior.
(b) Bright and dark contributions to
$-2\operatorname{Im}\Sigma^R(\omega_\mathrm{UP}^{\mathrm{pk}})$ for $N=4$
and $64$;
ensemble size uses the same colors as panel~(a), while dark and bright
channels are solid and dashed, respectively.  Here
$\omega_\mathrm{UP}^{\mathrm{pk}}$ is the center of the SCBA UP absorption
peak.
(c) Dark-channel contribution for $N=64$ at $T=0$, $100$, $200$, and
$300\,\mathrm K$.}
\label{fig:bright_dark_gap_scan}
\end{figure}

For $N>1$, SCBA predicts a UP-linewidth maximum about
$13.4\,\mathrm{meV}$ above the isolated-molecule value when the unperturbed
UP--dark gap is $10$--$20\,\mathrm{meV}$
[Fig.~\ref{fig:bright_dark_gap_scan}(a)].  We parameterize that gap, before
phonon self-energy corrections, as
\begin{equation}
\delta_{+D}\equiv\omega_+-\omega_0
=\frac{\Delta+\sqrt{\Delta^2+4G^2}}{2},
\qquad
\Delta=\omega_\mathrm{c}-\omega_0
=\delta_{+D}-\frac{G^2}{\delta_{+D}},
\label{eq:SI_gap_parameterization}
\end{equation}
where $G=\sqrt{N}g_\mathrm{c}$.  For the parameters used below,
$\omega_\mathrm{c}>0$ requires $\delta_{+D}>9.10\,\mathrm{meV}$; the
smaller-gap portion of the scan is a formal mathematical continuation rather
than a positive-frequency cavity configuration.  The tangent-grid calculation described in
Sec.~\ref{sec:SI_SCBA_numerics} approaches $\delta_{+D}=0$ directly.
The scan uses $G=136.2\,\mathrm{meV}$, $\omega_0=2.03\,\mathrm{eV}$,
$\kappa=44.15\,\mathrm{meV}$, and the same $300\,\mathrm K$
Drude--Lorentz bath as Fig.~\ref{fig:SI_detuning_additional}, with
$\lambda=30\,\mathrm{meV}$ and $\omega_f=24.8\,\mathrm{meV}$.
The corresponding isolated-molecule reference is
\begin{equation}
\Gamma_\mathrm{mol}^{\mathrm{SCBA}}=88.146\,\mathrm{meV},
\label{eq:SI_molecular_scba_width}
\end{equation}
obtained with $\eta=0.01\,\mathrm{meV}$.  It is $0.203\,\mathrm{meV}$ below
the $88.349\,\mathrm{meV}$ FFT-convolution check at
$\eta=0.1\,\mathrm{meV}$.  In the formal small-gap continuation, the UP
linewidth approaches the tangent-grid reference only below a
$1\,\mathrm{meV}$ gap.  The
$|C_{\mathrm{ex},+}|^2=0.97$ point in
Fig.~\ref{fig:SI_detuning_additional}(b), with
$\delta_{+D}\simeq24\,\mathrm{meV}$, remains in the excess-broadening
region.

We resolve the converged molecular self-energy into bright and dark internal
channels,
\begin{align}
\Sigma_B^R(\omega)&=\frac{1}{N}\Phi[G_{BB}^R](\omega),&
\Sigma_D^R(\omega)&=\frac{N-1}{N}\Phi[G_D^R](\omega),
\label{eq:SI_sigma_BD}\\
\Phi[G](\omega)&=\sum_\alpha c_\alpha^2
\left\{[\overline n(\omega_\alpha)+1]G^R(\omega-\omega_\alpha)
+\overline n(\omega_\alpha)G^R(\omega+\omega_\alpha)\right\}.
\notag
\end{align}
For each gap, the plotted channel values are evaluated at the center
$\omega_\mathrm{UP}^{\mathrm{pk}}$ of the SCBA UP absorption peak obtained by
the same peak analysis used for its FWHM; this dressed peak frequency is
distinct from the unperturbed $\omega_+$ used to define
$\delta_{+D}$.  At $N=64$ and $\delta_{+D}=20\,\mathrm{meV}$,
$-2\operatorname{Im}\Sigma_B^R=1.05\,\mathrm{meV}$ and
$-2\operatorname{Im}\Sigma_D^R=65.02\,\mathrm{meV}$ [panel~(b)].  The
linewidth excess and additional dark-channel damping have Pearson
correlations $r_\mathrm{P}\simeq0.90$--$0.91$ for $N=2$--64.  These
self-energy quantities resolve the internal propagator at the evaluated
frequency; the observed FWHM is extracted separately from the absorption
spectrum and is not their sum.

Cooling suppresses the dark contribution while leaving a finite zero-point
term [panel~(c)].  At the representative $20\,\mathrm{meV}$ gap, the grid,
range, broadening, convergence, and mixing variations in
Sec.~\ref{sec:SI_convergence} change the extracted FWHM by at most
$0.025\,\mathrm{meV}$; separate initialization tests remain stable at smaller
gaps.  Its absence in HEOM and
MF-HEOM suggests that omitted crossed or vertex-correction processes may be
important near degeneracy, but this channel decomposition does not identify
a unique missing diagram.

\FloatBarrier
\subsection{Collective-Limit Comparison with MF-HEOM}
\label{sec:SI_collective_limit}

\begin{figure}[htbp]
\centering
\includegraphics[width=0.92\linewidth]{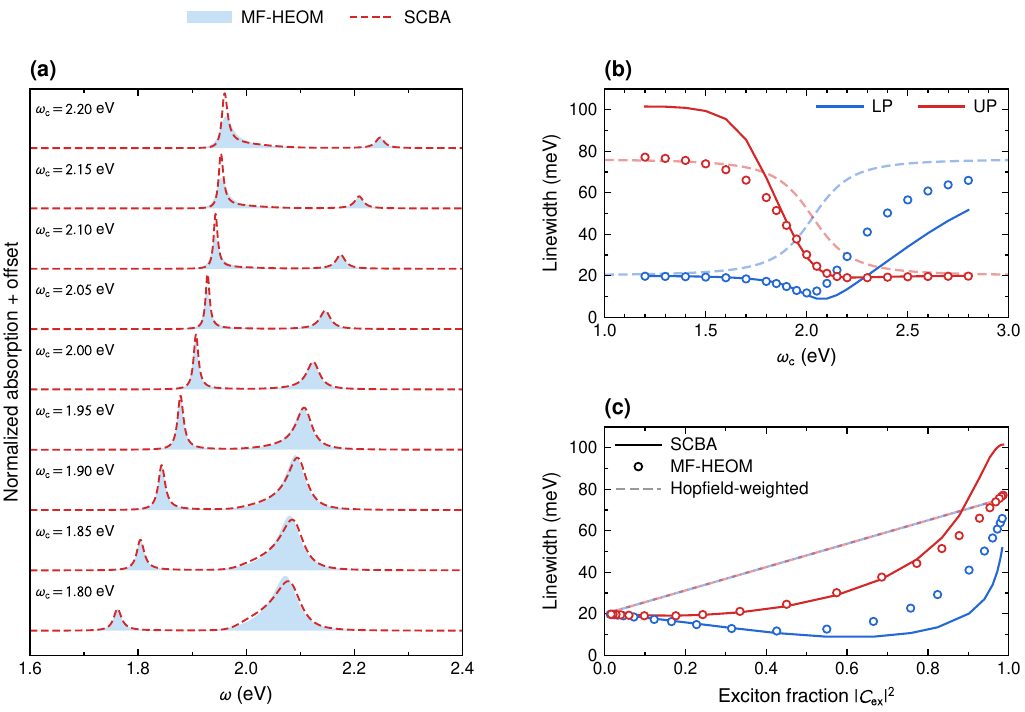}
\caption{Collective-limit comparison for $N=10^4$, a $300\,\mathrm K$
Drude--Lorentz bath with $\lambda=30\,\mathrm{meV}$ and
$\omega_f=24.8\,\mathrm{meV}$, $\kappa=20\,\mathrm{meV}$, and resonant
$\Omega_\mathrm{R}=0.2\,\mathrm{eV}$.  (a) Stacked absorption spectra for
$\omega_\mathrm{c}=1.80$--$2.20\,\mathrm{eV}$: MF-HEOM (blue shading) and
SCBA (red dashed).  Both profiles are normalized over
$1.6$--$2.4\,\mathrm{eV}$ and share a display factor at each cavity
frequency.  (b) UP (red) and LP (blue) FWHM in meV over the wider
cavity-frequency sweep: MF-HEOM (open circles), SCBA (solid), and
Hopfield-weighted dashed guides.  (c) The same linewidths versus exciton
fraction.}
\label{fig:collective_limit_comparison}
\end{figure}

Across the detuning sweep, SCBA follows the MF-HEOM polariton linewidths and
deviates most on the exciton-rich side, where the matter--phonon contribution
is strongest [Fig.~\ref{fig:collective_limit_comparison}].  We use $N=10^4$ to approximate the
collective limit.  This data set has cavity FWHM
$\kappa=20\,\mathrm{meV}$ and resonant collective Rabi splitting
$\Omega_\mathrm{R}=2\sqrt{N}g_\mathrm{c}=0.2\,\mathrm{eV}$, distinct from
the parameters of Fig.~\ref{fig:SI_detuning_additional}.  Panel~(a) displays
spectra for $\omega_\mathrm{c}=1.80$--$2.20\,\mathrm{eV}$; panels~(b,c)
use a wider $1.20$--$2.80\,\mathrm{eV}$ linewidth sweep.

The MF-HEOM spectrum is reconstructed as
$-\operatorname{Im}[P(\omega)/E(\omega)]$ from the propagated pulse in
Eq.~\ref{eq:SI_MF_pulse}.  Equation~\ref{eq:SI_MF_observable} establishes
the comparison with the same positive-frequency dipole spectral function
from SCBA.  Each spectrum is normalized by its area over
$1.6$--$2.4\,\mathrm{eV}$, and one common display factor is then applied to
the MF-HEOM--SCBA pair at each cavity frequency.  The finite-window area
checks are summarized with the other normalization diagnostics in
Sec.~\ref{sec:linewidth}.

\FloatBarrier

\clearpage
\bibliography{Ref}

%% file: Figures/Main/Fig1/Fig1.tex
% Editable TikZ source for main-text Fig. 1.
\begin{minipage}[t]{0.43\columnwidth}
\centering
\resizebox{\linewidth}{!}{%
\begin{tikzpicture}[
  every node/.style={font=\small},>=Stealth,
  paneltag/.style={anchor=north west,inner sep=2pt,font={\fontfamily{phv}\selectfont\LARGE}},
  mol/.style={circle,draw=blue!65!black,fill=blue!10,minimum size=14pt,inner sep=0pt,line width=0.8pt},
  gcoupl/.style={->,draw=blue!55!black,line width=0.85pt}]
\path[use as bounding box] (0,-0.05) rectangle (4.45,3.30);
\node[paneltag] at (0.05,2.85) {(a)};
\node[mol] at (1.90,2.15) {\tiny 1};
\node[mol] at (1.90,1.25) {\tiny 2};
\node[mol] at (1.90,0.25) {\tiny $N$};
\node[font=\normalsize] at (1.90,0.78) {$\vdots$};
\foreach \y in {2.15,1.25,0.25}{
  \draw[decorate,decoration={snake,amplitude=1.8pt,segment length=5pt},draw=gray!60,line width=0.75pt] (1.73,\y)--(1.10,\y);
  \draw[gray!55,line width=0.8pt] (1.08,{\y-0.13})--(1.08,{\y+0.13});}
\node[font=\normalsize,text=gray!50] at (1.08,0.78) {$\vdots$};
\node[font=\scriptsize,align=right,anchor=east] at (0.92,1.25) {local\\phonon\\bath};
\draw[gcoupl] (2.07,2.15)--(3.12,1.68);
\draw[gcoupl] (2.07,1.25)--(3.12,1.25);
\draw[gcoupl] (2.07,0.25)--(3.12,0.82);
\node[font=\scriptsize,text=blue!55!black,above=1pt] at (2.60,1.25) {$g_{\rm c}$};
\draw[draw=orange!60!black,fill=orange!5,line width=1pt] (3.60,1.25) circle (0.46cm);
\node[font=\scriptsize,text=orange!65!black,align=center] at (3.60,1.25) {cavity\\$\omega_{\rm c}$};
\end{tikzpicture}}
\end{minipage}\hfill
\begin{minipage}[t]{0.56\columnwidth}
\centering
\resizebox{\linewidth}{!}{%
\begin{tikzpicture}[
  every node/.style={font=\small},>=Stealth,
  paneltag/.style={anchor=north west,inner sep=2pt,font={\fontfamily{phv}\selectfont\LARGE}},
  level/.style={line width=1.5pt},
  phcoupl/.style={decorate,decoration={snake,amplitude=1.5pt,segment length=4pt,post length=1.5pt},orange!85!black,line width=0.65pt},
  phhead/.style={->,orange!85!black,line width=0.65pt},
  effcoupl/.style={decorate,decoration={snake,amplitude=2pt,segment length=5pt,post length=2pt},orange!85!black,line width=1.15pt},
  effhead/.style={->,orange!85!black,line width=1.15pt}]
\path[use as bounding box] (0,-0.05) rectangle (5.95,3.30);
\node[paneltag] at (0.05,2.85) {(b)};
\draw[level,blue!70!black] (1.25,2.16)--(5.20,2.16) node[anchor=west,font=\footnotesize,text=blue!70!black] {UP};
\draw[level,red!70!black] (1.25,0.14)--(5.20,0.14) node[anchor=west,font=\footnotesize,text=red!70!black] {LP};
\fill[gray!18] (2.00,0.82) rectangle (4.10,1.48);
\draw[gray!45,line width=0.55pt] (2.00,0.82) rectangle (4.10,1.48);
\foreach \y in {0.93,1.05,1.17,1.29,1.41}{\draw[gray!34,line width=0.35pt] (2.18,\y)--(3.92,\y);}
\node[font=\tiny,align=center] at (3.05,1.15) {$N{-}1$\\dark states};
\draw[phcoupl] (2.35,1.62)--(2.35,1.90); \draw[phhead] (2.35,1.90)--(2.35,2.00);
\draw[phcoupl] (2.58,2.00)--(2.58,1.72); \draw[phhead] (2.58,1.72)--(2.58,1.62);
\draw[phcoupl] (2.35,0.30)--(2.35,0.56); \draw[phhead] (2.35,0.56)--(2.35,0.66);
\draw[phcoupl] (2.58,0.66)--(2.58,0.40); \draw[phhead] (2.58,0.40)--(2.58,0.30);
\node[font=\scriptsize,text=orange!82!black,align=right,anchor=east] at (1.78,1.15) {phonon\\coupling};
\draw[effcoupl] (4.35,0.22)--(4.35,1.93); \draw[effhead] (4.35,1.93)--(4.35,2.08);
\draw[effcoupl] (4.60,2.08)--(4.60,0.37); \draw[effhead] (4.60,0.37)--(4.60,0.22);
\end{tikzpicture}}
\end{minipage}

\vspace{2.5mm}

\makebox[\columnwidth][l]{\fontfamily{phv}\selectfont\Large (c)}

\begin{minipage}[t]{0.455\columnwidth}
\centering
\resizebox{\linewidth}{!}{%
\begin{tikzpicture}[
  baseline=(current bounding box.center),
  scale=1.0,
  every node/.style={font=\small},
  every path/.style={line width=0.9pt},
  >=Stealth,
  midarrow/.style={
    postaction={decorate},
    decoration={markings, mark=at position 0.5 with {\arrow{Stealth[length=3.3mm]}}}
  },
  dressedarrow/.style={
    double,
    double distance=1.4pt,
    line width=0.9pt,
    postaction={decorate},
    decoration={markings, mark=at position 0.5 with {\arrow{Stealth[length=3.3mm]}}}
  },
  phonon/.style={
    decorate,
    decoration={snake, amplitude=5pt, segment length=11.7pt},
    line cap=round,
    line join=round
  }
]
\coordinate (i) at (0,0);
\coordinate (a) at (1.85,0);
\coordinate (b) at (4.45,0);
\coordinate (f) at (6.30,0);

\draw[midarrow] (i) -- (a);
\draw[midarrow] (a) -- (b);
\draw[midarrow] (b) -- (f);

\draw[phonon]
  (a) .. controls +(0.5,1.0) and +(-0.5,1.0) .. (b);

\fill (a) circle (1.1pt);
\fill (b) circle (1.1pt);

\node[below=4pt] at ($(a)!0.5!(b)$) {$G_0$};
\node[above=7pt] at ($(a)!0.5!(b)+(0,0.90)$) {$D_0$};
\end{tikzpicture}}
\end{minipage}\hspace{0.06\columnwidth}%
\begin{minipage}[t]{0.455\columnwidth}
\centering
\resizebox{\linewidth}{!}{%
\begin{tikzpicture}[
  baseline=(current bounding box.center),
  scale=1.0,
  every node/.style={font=\small},
  every path/.style={line width=0.9pt},
  >=Stealth,
  midarrow/.style={
    postaction={decorate},
    decoration={markings, mark=at position 0.5 with {\arrow{Stealth[length=3.3mm]}}}
  },
  dressedarrow/.style={
    double,
    double distance=1.4pt,
    line width=0.9pt,
    postaction={decorate},
    decoration={markings, mark=at position 0.5 with {\arrow{Stealth[length=3.3mm]}}}
  },
  phonon/.style={
    decorate,
    decoration={snake, amplitude=5pt, segment length=11.7pt},
    line cap=round,
    line join=round
  }
]
\coordinate (i) at (0,0);
\coordinate (a) at (1.85,0);
\coordinate (b) at (4.45,0);
\coordinate (f) at (6.30,0);

\draw[midarrow] (i) -- (a);
\draw[dressedarrow] (a) -- (b);
\draw[midarrow] (b) -- (f);

\draw[phonon]
  (a) .. controls +(0.5,1.0) and +(-0.5,1.0) .. (b);

\fill (a) circle (1.1pt);
\fill (b) circle (1.1pt);

\node[below=4pt] at ($(a)!0.5!(b)$) {$G$};
\node[above=7pt] at ($(a)!0.5!(b)+(0,0.90)$) {$D_0$};
\end{tikzpicture}}
\end{minipage}

%% file: Figures/SI/FigS1/FigS1.tex
% Editable TikZ source for SI Fig. S1.
\resizebox{\linewidth}{!}{%
\definecolor{phononblue}{RGB}{31,119,180}%
\begin{tikzpicture}[
  font=\sffamily\footnotesize,
  >=Stealth,
  exc/.style={draw=black,line width=0.8pt,
    postaction={decorate},
    decoration={markings,mark=at position 0.55 with
      {\arrow{Stealth[length=2.0mm,width=1.5mm]}}}},
  ph/.style={draw=phononblue,dashed,line width=0.9pt},
  crossph/.style={ph,preaction={draw=white,line width=2.2pt}},
  vtx/.style={circle,fill=black,inner sep=1.15pt},
  panel/.style={draw=gray!55,rounded corners=2pt,line width=0.45pt},
  ptitle/.style={anchor=west,font=\sffamily\bfseries\small}
]
\path[use as bounding box] (0,0) rectangle (15.2,8.4);
\draw[panel] (0.10,4.30) rectangle (7.45,8.30);
\draw[panel] (7.75,4.30) rectangle (15.10,8.30);
\draw[panel] (0.10,0.10) rectangle (7.45,4.05);
\draw[panel] (7.75,0.10) rectangle (15.10,4.05);

% (a) Fan diagram
\node[ptitle] at (0.35,7.95) {(a) Fan self-energy};
\node at (1.45,6.05) {$\boldsymbol\Sigma_{\mathrm{Fan}}[\mathbf G,D_0]=$};
\draw[exc] (2.80,6.00)--(6.80,6.00);
\draw[ph] (3.35,6.00) .. controls (3.65,7.25) and (5.75,7.25) .. (6.05,6.00);
\node[vtx] at (3.35,6.00) {};
\node[vtx] at (6.05,6.00) {};
\node[phononblue] at (4.70,7.15) {$D_0$};
\node at (4.70,5.62) {molecular $\mathbf G$};
\node at (3.78,4.70) {two vertices; dressed molecular line};

% (b) Dyson expansion of the self-consistent line
\node[ptitle] at (8.00,7.95) {(b) SCBA rainbow series};
\draw[exc] (8.10,5.95)--(9.75,5.95);
\draw[ph] (8.35,5.95) .. controls (8.50,6.85) and (9.35,6.85) .. (9.50,5.95);
\node[vtx] at (8.35,5.95) {};
\node[vtx] at (9.50,5.95) {};
\node at (10.02,6.25) {$+$};
\draw[exc] (10.25,5.95)--(12.10,5.95);
\draw[ph] (10.45,5.95) .. controls (10.65,7.25) and (11.70,7.25) .. (11.90,5.95);
\draw[ph] (10.75,5.95) .. controls (10.90,6.75) and (11.45,6.75) .. (11.60,5.95);
\foreach \x in {10.45,10.75,11.60,11.90}{\node[vtx] at (\x,5.95) {};}
\node at (12.38,6.25) {$+$};
\draw[exc] (12.60,5.95)--(14.25,5.95);
\draw[ph] (12.75,5.95) .. controls (12.90,7.35) and (13.95,7.35) .. (14.10,5.95);
\draw[ph] (12.98,5.95) .. controls (13.10,7.00) and (13.78,7.00) .. (13.90,5.95);
\draw[ph] (13.20,5.95) .. controls (13.30,6.70) and (13.58,6.70) .. (13.68,5.95);
\foreach \x in {12.75,12.98,13.20,13.68,13.90,14.10}{\node[vtx] at (\x,5.95) {};}
\node at (14.62,6.25) {$+\cdots$};
\node at (11.43,4.70) {included in SCBA};

% (c) crossed phonon diagram
\node[ptitle] at (0.35,3.70) {(c) Crossed/vertex diagram};
\draw[exc] (1.00,1.80)--(6.55,1.80);
\draw[ph] (1.80,1.80) .. controls (2.20,3.20) and (4.25,3.20) .. (4.65,1.80);
\draw[crossph] (2.85,1.80) .. controls (3.25,3.20) and (5.30,3.20) .. (5.70,1.80);
\foreach \x in {1.80,2.85,4.65,5.70}{\node[vtx] at (\x,1.80) {};}
\node[font=\bfseries] at (3.78,0.48) {beyond SCBA};

% (d) phonon polarization bubble
\node[ptitle] at (8.00,3.70) {(d) Phonon polarization};
\node at (8.55,2.20) {$\Pi^R=$};
\draw[ph] (9.05,2.20)--(9.65,2.20);
\draw[ph] (11.45,2.20)--(12.05,2.20);
\draw[exc] (9.65,2.20) .. controls (9.95,3.15) and (11.15,3.15) .. (11.45,2.20);
\draw[exc] (11.45,2.20) .. controls (11.15,1.25) and (9.95,1.25) .. (9.65,2.20);
\node[vtx] at (9.65,2.20) {};
\node[vtx] at (11.45,2.20) {};
\node at (12.40,2.20) {$=0$};
\node at (11.35,0.62) {$G^<=0\ \Longrightarrow\ D^R=D_0^R$};
\end{tikzpicture}%
}